\documentclass[aps,showpacs]{revtex4}

\usepackage{amssymb,amsthm}
\newcommand{\B}{{\bf B}}
\newcommand{\T}{{\bf T}}
\newcommand{\ui}{{\underline{i}}}
\newcommand{\uj}{{\underline{j}}}
\newcommand{\uk}{{\underline{k}}}
\newcommand{\ul}{{\underline{l}}}
\newcommand{\um}{{\underline{m}}}

\newcommand\be{\begin{eqnarray}}
\newcommand\ee{\end{eqnarray}}

\begin{document}

\title{Motion of a ``small body'' in non-metric gravity}

\author{Kirill Krasnov}
\affiliation{School of Mathematical Sciences, University of Nottingham, NG7 2RD, UK}
  
\date{v2: December 2008}

\begin{abstract} We describe ``small bodies'' in a non-metric gravity theory
previously studied by this author. The main dynamical field of the theory is a certain triple of 
two-forms rather than the metric, with only the spacetime conformal structure, not
metric, being canonically defined. The theory is obtained from general relativity (GR) in Plebanski
formulation by adding to the action a certain potential. Importantly, the modification
does not change the number of propagating degrees of freedom as compared to GR. We find 
that ``small bodies'' move along geodesics of a certain metric that is constructed with the help of a
new potential function that appears in the matter sector. We then use the ``small body'' 
results to formulate a prescription for coupling the theory to general 
stress-energy tensor. In its final formulation the theory takes an entirely standard form, with matter
propagating in a metric background and only the matter-gravity coupling and the gravitational dynamics being modified. 
This completes the construction of the theory and opens way to an analysis of its physical predictions.  
\end{abstract}


\maketitle

\section{Introduction}

In general relativity (GR) a test particle moves along a spacetime geodesic. 
The fact does not need to be added as a separate postulate of the theory. Indeed, general
relativity possesses a well-defined initial value formulation, so to
determine the evolution of a body one should just prescribe the initial data
for its gravitational field and read off the
trajectory from the resulting spacetime metric. For a ``small body'' this procedure gives
the geodesic motion, and the Bianchi identities satisfied by the Einstein
tensor are at the root of the derivation. A systematic procedure
that allows to derive not only the geodesic motion, but also 
the corrections to it (the so-called gravitational self-force) has been recently described 
in \cite{Gralla:2008fg}.

In \cite{Krasnov:2006du} the present author has described a large class of gravity 
theories that are based on two-forms rather than the metric. This class
contains general relativity (in Plebanski formulation \cite{Plebanski:1977zz}) 
and can be arrived at rather simply, see \cite{Krasnov:2008fm}, by dropping the 
simplicity constraints of Plebanski's theory. A theory from the class is specified by a certain 
``potential'' -- a scalar function of certain components of the two-form field. Exactly as GR,
the theory describes just two propagating degrees of freedom. The same class of gravity theories
has appeared much earlier in works of Bengtsson and Peldan under the
name of neighbours of GR, see e.g. \cite{Bengtsson:1990qg}. 
These authors' starting point (the pure connection formulation \cite{Capovilla:1991kx}) was, however, 
entirely different from that in \cite{Krasnov:2006du}, so the equivalence of models proposed
in \cite{Bengtsson:1990qg} to the theory described in
\cite{Krasnov:2006du} is not obvious and was pointed out in \cite{Bengtsson:2007zzd}.

In the version proposed by this author, the basic dynamical field of the theory is an ${\mathfrak su}(2)$ 
Lie algebra-valued two-form (complexified in the Lorentzian signature case). With the Lie algebra being
three-dimensional, the two-form field can be viewed as a triple of two-forms, and these can be 
declared to span the space of two-forms {\it self-dual} with respect to some metric. The
knowledge of which two-forms are self-dual can be shown \cite{Samuel} to determine the conformal structure 
of the metric uniquely. Thus, any theory based on ${\mathfrak su}(2)$ Lie algebra-valued
two-forms is naturally a theory of the conformal structure of spacetime.
However, it is by no means obvious which metric in the conformal class
so defined plays a physically distinguished role. Note that in Plebanski 
formulation of GR this problem does not arise as additional
simplicity constraints that are imposed on the two-form field
guarantee the existence of a preferred metric. It is not
even clear that there is {\it any} physically distinguished metric in the theory. 
Indeed, this would be a metric in which test particles move
along geodesics. To find whether there is such a metric, one
would need to describe how the ``usual'' matter couples
to the gravity theory in question. However, with the theory being
that of two-forms rather than the metric, this is an unsolved problem. The only case
that is currently understood is that of Yang-Mills fields which, being
classically conformally-invariant, do not help.

The goal of this paper is to develop the physical interpretation of
the theory \cite{Krasnov:2006du} by studying the motion of a ``small body''.
This allows us to sidestep the unsolved problem of coupling to 
generic matter and develop the physical interpretation of the theory remaining
entirely within its domain. We shall use the systematic procedure of \cite{Gralla:2008fg} that
needs only very little adaptation to the theory in question. 

Our main result is that there is a physically distinguished metric in the
theory \cite{Krasnov:2006du} along whose geodesics test particles move. 
However, we find that it is the matter itself that supplies the
conformal factor that determines this metric. Thus, we shall see that the
coupling of matter to the gravity theory in question is characterized
by a certain ``mass'' function, and the metric in which particles
move along geodesics is obtained by choosing the conformal
factor so that this function is a constant. If the theory is to preserve the weak 
equivalence principle the mass function of all material bodies must be the same. This
requirement introduces a certain universal potential function, see the main text. A very similar
function appears on the gravitational side, and the theory is thus completely
specified by prescribing the gravity and matter side potentials. 

Having obtained an expression for the stress-energy ``tensor'' of a small body, 
it is not hard to extend it to a description of how the general stress-energy tensor of matter
couples to our gravity theory. We give such a description, thus
completing the construction of the theory and making the study of its physical 
predictions possible. We would like to emphasize at the outset that, in spite of
the metric appearing in this theory only indirectly,
the final formulation of the theory is entirely standard: one has
usual matter fields moving in a metric background.
Only the dynamics of gravity, as well as its coupling to matter are
modified. However, unlike all previous modification schemes considered
in the literature, the theory in question modifies both vacuum and non-vacuum GR 
without adding to it any new propagating degrees of freedom.
We would like to stress that this feature of the theories considered here is quite 
striking, for it is a rather common belief that the only way
to modify Einstein's theory is to add to it new propagating modes.

The organization of this paper is as follows. In the next Section
we review the Plebanski formalism for general relativity. We also
describe how matter (e.g. ideal fluid) can be coupled to gravity in
this formulation. In Section \ref{sec:mod} we describe the modified
gravity theories \cite{Krasnov:2006du}. We extend this description to the 
non-vacuum case in Section \ref{sec:non-vac}. In Section \ref{sec:stress-energy}
we obtain an expression for the stress-energy-momentum two-form of
a ``small body'' and use Bianchi identities to determine its
motion in Section \ref{sec:motion}. An interpretation of the
equations we obtain is contained in Section \ref{sec:interp}.
We describe how a general stress-energy tensor is coupled to
the gravity theory in question in section \ref{sec:gen}.
Section \ref{sec:recep} gives a metric formulation of the theory 
that is most useful for practical applications. 
We conclude with a discussion.

\section{Plebanski formalism}
\label{sec:Pleb}

The aim of this Section is to review the Plebanski formulation of
general relativity. As we have already mentioned, in this formulation
Einstein's gravity becomes a theory of two-forms rather than the metric.
Plebanski's formalism uses in a deep way the notions of self-duality
on two forms (as determined by the spacetime metric) and it is the
process of ``abstracting'' this notion from the underlying metric
that allows for a deep reformulation of Einstein's theory. The
original Plebanski's paper \cite{Plebanski:1977zz} used spinor
notations and is not very transparent for a reader who is not
familiar with spinor techniques. An excellent exposition of the theory
is also available in \cite{Capovilla:1991qb}, where the problem
of coupling to matter is discussed as well. 

In this paper, to make it more easy to follow, we will try to avoid using spinors as 
hard as possible, only resorting to spinor techniques when they simplify
computations. All such spinor calculations are banned to the Appendix.

We have also decided to make the exposition of the Plebanski formalism
as concrete as possible, so in this section we present it as a concrete
recipe for deriving Einstein equations. However, before we give such
an explicit description, let us state the main ideas abstractly.

Plebanski theory introduces a (complexified in Lorentzian signature) SO(3) 
vector bundle $V$ over the spacetime $M$, which can be
referred to as the self-dual bundle, and a two-form field $B^i, i=1,2,3$ taking values in $V$.
The triple of two forms $B^i, i=1,2,3$ encodes information about the metric $g$ on $M$ via the requirement 
that $B^i, i=1,2,3$ are self-dual two-forms with respect to $g$. Indeed, the triple $B^i$ spans a
3-dimensional subspace in the space of all two-forms, and declaring this to be the
subspace of self-dual two forms defines the notion of Hodge duality on two forms, which, in turn, 
can be shown \cite{Samuel} to uniquely determine the conformal class of the metric.  
However, a general triple $B^i$ of two-forms contains too many components as compared to a metric. 
Indeed, it needs $3\times 6$ numbers to be specified, while a metric has only 10 components. 
To remedy this, Plebanski imposes the following ``metricity'' (or simplicity) conditions:
\be\label{metr}
B^i \wedge B^j \sim \delta^{ij},
\ee
which give 5 equations on the two-form field (the trace of this equation gives the proportionality
coefficient and is an identity). This brings the number of components
in $B^i$ down to 13, which is the required 10 components describing the metric, plus
3 gauge components related to availability of SO(3) gauge transformations. The volume
form of this metric is then given by $(i/3)(B^i\wedge B^i)$. With the two-form
field $B^i$ being complex, one further needs 10 conditions that guarantee
that the metric obtained is real Lorentzian. These conditions are given below
in (\ref{B-conds}).

Thus, supplemented by the metricity conditions (\ref{metr}) the two-form field contains
just the right amount of information to describe a metric. One now needs a second order
differential equation on $B^i$. To obtain this one notices that there is a unique
connection $A^i$ satisfying:
\be\label{comp}
D_A B^i =0,
\ee
where $D_A B^i:=dB^i + \epsilon^{ijk} A^j\wedge B^k$.
Indeed, this gives $4\times 3$ algebraic equations for $4\times 3$ components of $A^i$,
which fixes it uniquely, provided a certain non-degeneracy conditions for $B^i$ are
satisfied. We shall refer to this connection as $B$-compatible and denote it by $A_B$. When $B^i$ satisfies
the metricity conditions (\ref{metr}) the $B$-compatible connection turns out to
be equal to the self-dual part of the metric-compatible one. One can now compute the curvature 
$F^i:=dA^i + (1/2)\epsilon^{ijk}A^j\wedge A^k$ of the $B$-compatible connection.
A natural second order equation on $B^i$ is obtained by requiring that the
curvature $F^i$ is ``proportional'' to the two-form field $B^i$. However, it is natural to 
allow for the ``proportionality coefficient'' to be an ``internal'' tensor:
\be\label{eq-Pleb}
F^i = \Phi^{ij} B^j,
\ee 
where the quantities $\Phi^{ij}$ are at this stage arbitrary. It can then be shown that the 
``internal'' tensor $\Phi^{ij}$ must be symmetric and that its trace part must be a constant:
\be
\Phi^{ij}=\Psi^{ij} + \frac{1}{3} \delta^{ij} \Lambda, \qquad \Psi^{ij}=\Psi^{(ij)}, 
\qquad \Psi^{ij}\delta_{ij}=0, \qquad \Lambda=const.
\ee
The relation (\ref{eq-Pleb}) then becomes a set of 18 equations for 13 components of the $B$-field as well
as for 5 (traceless symmetric) undetermined components $\Psi^{ij}$ of $\Phi^{ij}$. To see that
(\ref{eq-Pleb}) is equivalent to Einstein equations one notices that it states
that the curvature of the self-dual part of the spin connection is self-dual as
a two-form. This is known to be equivalent to the Einstein condition. 

Finally, let us note that all equations of Plebanski theory can be obtained
as Euler-Lagrange equations for the following action:
\be\label{action-Pleb}
S[B,A,\Psi]=\int B^i\wedge F^i(A) - \frac{1}{2}\left( \Psi^{ij} + \frac{\Lambda}{3}\delta^{ij}
\right) B^i\wedge B^j.
\ee
Indeed, the variation with respect to the traceless tensor $\Psi^{ij}$ gives (\ref{metr}), 
variation with respect to the connection gives (\ref{comp}), while variation with
respect to the two-form field gives the main dynamical equation (\ref{eq-Pleb}).

To make the above description little less abstract let us reformulate it
as a concrete recipe for writing down Einstein equations once a spacetime
metric is given. The starting point of Plebanski method of deriving Einstein equations 
is the same as in the tetrad method: one has to find a
suitable tetrad. For a diagonal metric there is no ambiguity, but for
a non-diagonal one it is possible to use the available freedom of Lorentz rotations
to bring the tetrad to a convenient form. Thus, we assume that we have
found a convenient collection of one-forms $e^I, i=0,1,2,3$ so that the metric is:
\be\label{metric}
ds^2 = e^I \otimes e^J \eta_{IJ},
\ee
where $\eta_{IJ}$ is the Minkowski metric. 

The second step is to form a set of three
two-forms $B^i, i=1,2,3$ which are self-dual with respect to the given metric, and
satisfy: 
\be\label{B-conds}
B^i\wedge B^j\sim \delta^{ij}, \qquad B^i\wedge(B^j)^*=0, \qquad {\rm Re}(B^i\wedge B^i)=0,
\ee
where $(B^i)^*$ are the complex conjugate two-forms. This task is easy if one
has a tetrad in one's disposal, with a possible solution being:
\be\label{B-tetr}
B^1=i e^0\wedge e^1 - e^2\wedge e^3, \qquad
B^2=i e^0\wedge e^2 - e^3\wedge e^1, \qquad
B^3=i e^0\wedge e^3 - e^1\wedge e^2.
\ee
It is easy to see that all the required conditions (\ref{B-conds}) are satisfied,
and that the above two-forms are indeed self-dual with respect to (\ref{metr}).
It can also be shown that the converse is true: given a triple of two-forms 
$B^i$ satisfying (\ref{B-conds}) there is a unique real metric with respect
to which the two-forms $B^i$ are self-dual.

The third step is to find an $\mathfrak{su}(2)$ connection $A^i$ that is
``compatible'' with the above set $B^i$ of two-forms, in the sense that
the covariant derivative of $B^i$ with respect to $A^i$ is zero:
$D_A B^i=dB^i + \epsilon^{ijk} A^j\wedge B^k=0$. To obtain
such a connection one has to solve the following system of linear
algebraic equations for the components of the connection:
\be\label{B-compat}
dB^1+A^2\wedge B^3 - A^3\wedge B^2=0, \qquad
dB^2+A^3\wedge B^1 - A^1\wedge B^3=0, \qquad
dB^3 + A^1\wedge B^2 - A^2\wedge B^1=0.
\ee
In doing so we first have to take the exterior derivative of the three
two forms $B^i$, and then write down the above system of $3\times 4$
equations for the connection components $a^i_I: A^i = a^i_I \theta^I$.
This is an exercise in algebra, of roughly the same
degree of complexity as one that arises in the determination of the
rotation coefficients in the tetrad-based approach. What simplifies
the game somewhat is that in the tetrad-based approach, at least in 
principle, one has $4\times 6$ equations to write, for the same number 
of the rotation coefficients, while in the case of Plebanski formulation
there is just half that number. No information is lost, however, as
all quantities in the Plebanski case are complex. In practice, 
for a given metric (usually possessing some symmetry properties), 
most of the connection coefficients are zero by symmetries both in the tetrad and
the two-form cases, so the amount of work one has to do to find $A^i$
is only a little less than in the tetrad-based scheme.
As in the tetrad-based scheme it is much easier to
verify a conjectural solution than find one, as the latter involves some
guesswork on which of the components are zero. Finding the connection
is the most laborious part of the computation.  

The fourth step is to compute the curvature two-form 
$F^i= dA^i+(1/2)\epsilon^{ijk}A^j\wedge A^k$. In components:
\be\label{curv}
F^1 = dA^1 + A^2\wedge A^3, \qquad F^2 = dA^2 + A^3\wedge A^1, \qquad
F^3=dA^3+A^1\wedge A^2.
\ee
This is a simple exercise in differentiation. The step of computing $F^i$ should be 
compared to the curvature computation in the tetrad-based approach. In that case one needs 
to compute six two-forms, as compared to only three in the Plebanski case.

The fifth step is to write all the two-forms that appear in $F^i$ in terms of the six 
basic two-forms that are used in (\ref{B-tetr}). This may involve some
algebra in the case the metric is non-diagonal. This step is exactly
the same as in the tetrad-based scheme, when one writes the curvature
components in terms of the basic tetrad two-forms (before the Ricci tensor 
can be found by contracting a pair of indices). 

The final step is to replace the tetrad two-forms appearing in the result for
$F^i$ by their expressions in terms of $B^i$ and $(B^i)^*$. It turns out to be
more convenient to use the two-forms $\bar{B}^i=-(B^i)^*$. We have:
\be\nonumber
e^0\wedge e^1 = \frac{1}{2i} (B^1 + \bar{B}^1), \qquad 
e^2\wedge e^3 = \frac{1}{2} (\bar{B}^1-B^1), \\ \label{tetr-B}
e^0\wedge e^2 = \frac{1}{2i} (B^2 + \bar{B}^2), \qquad 
e^3\wedge e^1 = \frac{1}{2} (\bar{B}^2-B^2), \\ \nonumber
e^0\wedge e^3 = \frac{1}{2i} (B^3 + \bar{B}^2), \qquad 
e^1\wedge e^2 = \frac{1}{2} (\bar{B}^3-B^3).
\ee
This last step does not have a direct analog in the tetrad-based method.

This is it! Once an expression for the curvature $F^i$ in terms
of the self- and anti-self dual two-forms is found, one can immediately
write down the Einstein equations. Indeed, we have obtained the curvature
in the form:
\be\label{M-N}
F^i = M^{ij} B^j +N^{ij} \bar{B}^j,
\ee
where $M^{ij}, N^{ij}$ are some matrices built from the components
of the metric and their first and second derivatives. The Plebanski field equations read:
\be\label{equation-Pleb}
F^i = \left(\Psi^{ij} + \frac{1}{3} \delta^{ij} \Lambda \right) B^j + 4\pi G T^i,
\ee
where $\Psi^{ij}$ is a traceless matrix, $\Lambda$ is the cosmological constant 
and $T^i$ is the ``stress-energy-momentum two-form'', 
see below, zero in vacuum. Thus, the vacuum Einstein equations are simply:
\be\label{P}
{\rm Tr} M = \Lambda, \qquad N^{ij} = 0,
\ee
which gives ten equations, as it should. As a bonus, we also obtain the components
of the (self-dual part of the) Weyl curvature tensor. Indeed, we have $\Psi^{ij}=(M^{ij})_{(tf)}$, with
$(tf)$ standing for the tracefree part. 

Let us also write the general non-vacuum equations. To this end we need a general
expression for the ``stress-energy-momentum two-form'' in terms of the quantities characterizing
the gravitating matter. A general Lie-algebra valued two-form admits an 
expansion of the type (\ref{M-N}), thus giving rise to $9+9=18$ components. However, in 
general relativity the ``right-hand-side'' of Einstein equations -- the matter stress-energy-momentum 
tensor -- has ten components. Thus, the stress-energy-momentum two form of Plebanski formalism
cannot be a general Lie-algebra valued two-form. It needs to satisfy:
\be\label{T-conds}
T^i \wedge B^j \sim \delta^{ij},
\ee
which should be compared to the first of the conditions in (\ref{B-conds}), plus 
certain reality conditions, see below. The conditions (\ref{T-conds}) imply that $T^i$ has 
the following structure:
\be\label{T-general}
T^i = \frac{1}{6} T B^i - \frac{1}{2} T^{ij} \bar{B}^j,
\ee 
where the numerical prefactors are added for future convenience, and
$T, T^{ij}$ are an arbitrary scalar and $3\times 3$ ``internal'' tensor correspondingly.
This gives in total $10$ components in $T^i$ as should be expected from
a quantity representing the usual stress-energy tensor $T_{\mu\nu}$. The
interpretation of the components in (\ref{T-general}) is as follows. The scalar
$T$ is just the trace $T=g^{\mu\nu}T_{\mu\nu}$
of the usual stress-energy tensor. The $3\times 3$ tensor $T^{ij}$ can in turn be
decomposed into its symmetric and anti-symmetric parts:
\be\label{T-interpr}
T^{ij} = \sigma^{ij} + i\epsilon^{ijk} u^k,
\ee
and takes care of the traceless $T_{\mu\nu}-(1/4)g_{\mu\nu} T$ part of $T_{\mu\nu}$.
The symmetric part $\sigma^{ij}$ turns out to have the meaning of the spatial
components of the stress-energy tensor, i.e. characterize the stress of matter.
The anti-symmetric part $u^i$ gives the momentum vector. The quantities $T, \sigma^{ij}, u^i$
are all required to be real. Thus, the symmetric part of $T^{ij}$ is required to be real while
its anti-symmetric part is purely imaginary. 

The non-vacuum Einstein equations then take the form:
\be
{\rm Tr} M = \Lambda + 2\pi G T, \qquad N^{ij} = - 2\pi G T^{ij}.
\ee

For many applications one is interested in the simplest type of matter -- that
given by the perfect fluid. For the perfect fluid only the trace part of the stress tensor
is different from zero: $\sigma^{ij}\sim\delta^{ij}$. It is thus completely characterized by
its energy $\rho$ and pressure $P$ densities and the momentum vector $u^i$. 
The corresponding stress-energy-momentum two-form is:
\be\label{T-fluid}
T^i_{fluid} = \frac{1}{6} (\rho- 3P) B^i 
- \frac{1}{2} \left( (\rho+P) \delta^{ij} + i\epsilon^{ijk} u^k\right)\bar{B}^j.
\ee

\section{Modified gravity: The vacuum case}
\label{sec:mod}

We will start our presentation of the theory \cite{Krasnov:2006du} by describing 
the vacuum case. We follow closely a recent description \cite{Krasnov:2008fm}.

The class of theories in question can be obtained by relaxing the simplicity
conditions (\ref{metr}) of the Plebanski theory. Thus, the main idea is to
allow all the components of the two-form field $B^i$ to become dynamical. Recall that a general
two-form field $B^i$ determines a conformal structure of spacetime by requiring that
the triple $B^i, i=1,2,3$ is self-dual in this conformal structure. Thus, we are
going to obtain a theory in which the spacetime conformal structure, not the metric
becomes the main dynamical object. 

How can one get natural field equations describing the dynamics of $B^i$? As in Plebanski formulation
of GR one first computes the $B$-compatible connection $A_B$ and then its curvature 
$F(A_B)$. As before, it is natural and instructive to decompose the curvature into the basis of
two-forms $B^i$ and $\bar{B}^i$:
\be\label{F-decomp}
F^i(A_B) = M^{ij} B^j + N^{ij} \bar{B}^j.
\ee
The most natural field equations are the same as in the Plebanski case (\ref{eq-Pleb}). Thus, we
 require the curvature of the connection $A_B$ to be purely self-dual:
\be\label{field-eqs-vac}
F^i(A_B)=\Phi^{ij} B^j \qquad \Longleftrightarrow \qquad M^{ij}=\Phi^{ij},
\quad N^{ij} = 0,
\ee
where $\Phi^{ij}$ is some purely gravitational tensor to be described below.
The system of equations (\ref{field-eqs-vac}) gives us 18 equations for the
18 components of the two-form field $B^i$. However, it also contains the
so-far unspecified functions $\Phi^{ij}$ and so is not complete. Note that
in the GR case we have exactly the same system of 18 equations, but in that
case for 18-5 quantities $B^i$ (the two-form field $B^i$ modulo the conditions
$B^i\wedge B^j\sim\delta^{ij}$). In addition the trace part of $\Phi^{ij}$
is either zero (no cosmological constant case) or constant (the cosmological 
constant), and is thus not an unknown field. The system of 18 equations
is thus that for 13 components of $B^i$ and the remaining 5 components
of $\Phi^{ij}$.

In the general case the system of equations (\ref{field-eqs-vac}) can be completed 
by considering analogs of ``Bianchi'' identities.
Thus, we note that the components of $M^{ij}, N^{ij}$ in (\ref{F-decomp}) are
not independent. Indeed, we have the following Bianchi identity:
\be
D_{A_B} F(A_B) = 0.
\ee
This gives:
\be\label{bianchi-1}
(D_{A_B} M^{ij}) \wedge B^j + (D_{A_B} N^{ij} \bar{B}^j)=0.
\ee
Another important identity is obtained by using the compatibility equations
$D_{A_B} B^i=0$. Taking another covariant derivative and using the
definition of the curvature we get:
\be\label{bianchi-2}
\epsilon^{ijk} F^j(A_B) \wedge B^k = 0 \Longleftrightarrow \epsilon^{ijk} M^{jl} B^l\wedge B^k =0.
\ee
This last equation can be conveniently interpreted as follows. Let us define a conformal ``internal'' metric:
\be\label{h-metric}
B^i\wedge B^j \sim h^{ij}.
\ee
Then (\ref{bianchi-2}) can be rewritten as:
\be
\epsilon^{ijk} M^{jl} h^{lk} =0.
\ee

Let us now also introduce an action principle that leads to (\ref{field-eqs-vac}) as
Euler-Lagrange equations. This is easy to write, we have:
\be\label{action-full-vac}
S[B,A,\Phi]=\int B^i\wedge F^i(A) - \frac{1}{2}\Phi^{ij} B^i\wedge B^j.
\ee
Varying this with respect to $A^i$ we get $D_A B^i=0$, which allows to solve for $A$ in
terms of $B$, varying the action with respect to $B^i$ we get (\ref{field-eqs-vac}).
We also note that only the symmetric part of the field $\Phi^{ij}$ enters the
action, so it is necessary to assume that $\Phi^{ij}$ in (\ref{field-eqs-vac}) is symmetric.

It remains to clarify the meaning of the variation with respect to $\Phi^{ij}$. To
these end we shall use the Bianchi identities (\ref{bianchi-1}), (\ref{bianchi-2}).
Using (\ref{bianchi-1}) and field equations (\ref{field-eqs-vac}) we see that we
must have:
\be
D_{A_B} \Phi^{ij}\wedge B^j =0.
\ee
Let us multiply this equation by the one-form $\iota_\xi B^i$ and sum over $i$. Here
$\xi$ is an arbitrary vector field and  $\iota_\xi B^i$ is one-form with components
$(\iota_\xi B^i)_\mu:=\xi^\alpha B_{\alpha\mu}^i$. However, for any vector field $\xi$ we have:
\be\label{key-ident}
\iota_\xi B^{(i} \wedge B^{j)} = \frac{1}{2} \iota_\xi (B^i \wedge B^j) \sim h^{ij},
\ee
where $h^{ij}$ is the internal metric introduced above. This gives us the following equation:
\be
h^{ij} D_{A_B} \Phi^{ij} = 0.
\ee
Now, using the symmetry of $h^{ij}$ we can rewrite this equation as 
$h^{ij}(d\Phi^{ij} + 2\epsilon^{ikl} A^k \Phi^{lj})=0$. However,
the other Bianchi identity (\ref{bianchi-2}) together with the field equation
$M^{ij}=\Phi^{ij}$ implies $\epsilon^{kli} \Phi^{lj} h^{ji} =0$ and so we must have:
\be\label{h-Phi-ident}
h^{ij} d\Phi^{ij}=0.
\ee

The identity (\ref{h-Phi-ident}) implies that the quantities $h^{ij}$ and $\Phi^{ij}$ are
not independent. This can be seen quite clearly by considering the last term in 
the action (\ref{action-full-vac}). Using the introduced above tensor $h^{ij}$ we
can write the integrand as $V:=h^{ij} \Phi^{ij}$ times some volume form. We then get:
\be\label{dV}
dV=\Phi^{ij} dh^{ij} + h^{ij} d\Phi^{ij} =  \Phi^{ij} dh^{ij},
\ee
where we have used (\ref{h-Phi-ident}). This means that (i) the last term
in the action is only a function of the $h^{ij}$ components of the two-form
field $B^i$; (ii) the quantities $\Phi^{ij}$ are expressible through
$h^{ij}$ and are given by:
\be\label{Phi-h}
\Phi^{ij} = \frac{\partial V(h)}{\partial h^{ij}}.
\ee
Below we shall characterize the ``potential'' $V(h)$ in more details.
For now let us note that having expressed the unknown functions $\Phi^{ij}$ in
terms of the the components of the two-form field $B^i$ we have closed the
system of equations (\ref{field-eqs-vac}), as it is now a system of 18 equations
for 18 unknowns - components of the $B^i$ field.

To understand the structure of the potential $V(h)$ it is convenient to
parametrize the ``internal'' metric $h^{ij}$ by its trace and the traceless part:
\be\label{H-h}
h^{ij} = \frac{1}{3} {\rm Tr}(h) \left(\delta^{ij} + H^{ij}\right),
\ee
where $H^{ij}$ is tracefree. It is then easy to see that for any function 
$f(h^{ij})=f({\rm Tr}(h),H^{ij})$
\be\label{der-ident}
\frac{\partial f}{\partial h^{ij}} = \left( \frac{\partial f}{\partial {\rm Tr}(h)}
- \frac{\partial f}{\partial H^{kl}} \frac{H^{kl}}{{\rm Tr}(h)} \right) \delta^{ij}
+ \frac{3}{{\rm Tr}(h)} \frac{\partial f}{\partial H^{ij}}.
\ee
In particular, we have
\be
\frac{\partial f}{\partial h^{ij}} h^{ij} = \frac{\partial f}{\partial {\rm Tr}(h)} {\rm Tr}(h).
\ee
Thus, one has:
\be
V=\Phi^{ij} h^{ij} = {\rm Tr}(h) \frac{\partial V}{\partial {\rm Tr}(h)},
\ee
where $\Phi^{ij}$ is given by (\ref{Phi-h}). Thus, we learn that the 
potential must be a homogeneous function of order one in its argument ${\rm Tr}(h)$:
\be\label{V}
V(h) =  \Lambda \frac{{\rm Tr}(h)}{3} U(H),
\ee
where $U$ is a dimensionless function that only depends on the tracefree part $H^{ij}$ 
of $h^{ij}$, and is normalized so that $U(0)=0$, that is:
\be\label{U-exp}
U(H)=1+ \frac{\alpha}{2} {\rm Tr}(H^2) + O(H^3),
\ee
where $\alpha$ is some dimensionless parameter.
The quantity $\Lambda$ is a constant of dimensions $1/L^2$ that needs to
be introduced to give $V(h)$ the correct dimensions. Below it will be
identified with the cosmological constant. Using our definition of the ``internal'' metric $h^{ij}$ 
we can write:
\be
B^i \wedge B^j = \frac{h^{ij}}{{\rm Tr}(h)} (B^k\wedge B^k),
\ee
which of course defines $h^{ij}$ only up to a conformal factor.
One can now write down the action (\ref{action-full-vac}) as a functional of only the two-form 
and the connection fields:
\be\label{action-pot}
S[B,A]=\int B^i\wedge F^i(A) - \Lambda \frac{U(H)}{6} B^i\wedge B^i,
\ee
where $H^{ij}$ is defined as the traceless part of the internal metric $h^{ij}$,
and is independent of the conformal freedom present in the definition of $h^{ij}$.
Note that this action is an off-shell one, that is it can be varied with respect
to the dynamical fields $B^i, A^i$ to obtain field equations. Alternatively, one
can work with a version that uses extra ``Lagrange multiplier'' fields, see below.

Note that the dimensionfull constant $\Lambda$ here should be
identified with the cosmological constant of our theory. Indeed, one
can, e.g., consider the metric describing a homogeneous isotropic Universe. In such a 
Universe $H=0$ by symmetries, and so the metric evolves exactly like in 
general relativity with the cosmological constant $\Lambda$. In other words,
a solution of our theory describing a homogeneous isotropic Universe is the same as in 
GR with cosmological constant $\Lambda$. Thus,
to define the modified gravity theory in question one only needs to specify a dimensionless
function $U(H)$ of a dimensionless traceless tensor $H^{ij}$. 
All physical dimensionfull parameters present in the theory are 
as in general relativity. Let us remark that we could have
included $\Lambda$ into the definition of the potential $U(H)$
and thus made it dimensionfull. However, as we shall see below from 
matter coupling considerations, it is more convenient to make the potential function 
dimensionless, for one can then use similar potential functions in both the
gravity and the matter sectors.

In terms of the introduced potential $U(H)$ the main set
(\ref{field-eqs-vac}) of field equations becomes:
\be\label{field-eqs-pot}
F^i(A_B) = \Lambda \left( \frac{\partial U}{\partial H^{ij}} +
\frac{1}{3} \delta^{ij} \tilde{U} \right) B^j,
\ee
where we have introduced the Legendre transform $\tilde{U}$ of the potential $U$:
\be\label{Legendre}
\tilde{U}:= U - \frac{\partial U}{\partial H^{kl}} H^{kl}.
\ee
The function $\tilde{U}$ can be viewed as either that of $H^{ij}$ or of the
quantities:
\be
\Psi^{ij}/\Lambda := \frac{\partial U}{\partial H^{ij}}.
\ee
The theory with an arbitrary ``cosmological function'' 
$\Lambda(\Psi):=\Lambda \tilde{U}(\Psi/\Lambda)$ defined by the Lagrangian
\be\label{action-psi}
S[B,A]=\int B^i\wedge F^i(A) - \frac{1}{2}\left( \Psi^{ij} - \frac{1}{3}\Lambda(\Psi) \delta^{ij} \right) 
B^i\wedge B^i,
\ee
is that of the original paper \cite{Krasnov:2006du}. Field equations (\ref{field-eqs-pot})
are most easily obtained precisely in this ``Lagrange multiplier'' formulation. However, the
viewpoint suggested by (\ref{action-pot}), namely that of the gravity theory
being the BF theory (the first term in (\ref{action-pot})) plus a potential term 
for the $H^{ij}$ components of the two-form field will be more convenient for our purposes
here. 

To summarize, we have seen that the condition $B^i\wedge B^j\sim \delta^{ij}$ of Plebanski
formulation of GR can be relaxed and how the Bianchi identities still lead
(in a unique way) to a consistent theory. Note that what one obtains is  
a class of gravity theories rather than one theory, for a theory is now specified by 
a choice of the dimensionless ``potential'' function $U(H)$ of the components $H^{ij}$
of the two-form field $B^i$. The potential can be completely arbitrary. One can obtain
back general relativity (in Plebanski formulation) by making the potential function
$U(H)$ infinitely steep so that the quantities $H^{ij}$ are set to zero. However, if one
sets $U(H)=1$ one obtains a topological theory - the so-called BF theory
with a cosmological constant.

Having achieved a formulation of the theory in vacuum it is
very important to continue to develop the theory and allow for a 
non-trivial right hand side of our equations - for matter to be present.
Indeed, pure gravity is only of academic interest, and the real world gravity is
both produced and felt by material bodies.

\section{Modified gravity: Non-vacuum case}
\label{sec:non-vac}

The vacuum field equations (\ref{field-eqs-vac}) of modified gravity were exactly
the same as those (\ref{eq-Pleb}) of the vacuum Plebanski theory. As in Plebanski case,
it is natural to describe the effect of matter on the modified gravity ``geometry'' by
allowing a non-zero $T^i$ to be present on the right hand side of (\ref{field-eqs-vac}). 
This is in the spirit of Einstein equations, where the stress-energy-momentum of matter
appears on the right hand side of an equation for the curvature and thus affects the geometry.
This in turn implies that the stress-energy-momentum of matter should satisfy some conservation 
laws which to a large effect determine its motion in a given background. 

Thus, we shall keep the field equations (\ref{eq-Pleb}) as our main dynamical equations
even in the case of non-zero $T^i$. However, now that we have removed the condition
$B^i\wedge B^j\sim\delta^{ij}$ it no longer consistent to impose the condition
$T^i\wedge B^j\sim\delta^{ij}$ either. In general, the matter ``stress-energy-momentum'' two-form $T^i$ 
will have all components:
\be\label{T-decomp}
T^i = \frac{1}{2} Q^{ij} B^j - \frac{1}{2} T^{ij} \bar{B}^j,
\ee
where the interpretation of components $T^{ij}$ is similar to that in GR,
see (\ref{T-interpr}), and $Q^{ij}$ are some new ``internal'' components of $T^i$.
We note that, in the case $Q^{ij}\sim\delta^{ij}$, the trace $T$ of the usual GR stress-energy-momentum 
tensor is just a multiple of the trace ${\rm Tr}(Q)$.
The quantity $Q^{ij}$, by analogy with $T^{ij}$ can be referred to as
the ``internal'' stress-momentum of matter. Let us decompose $Q^{ij}$
into its symmetric and anti-symmetric parts:
\be
Q^{ij} = \chi^{ij} + i\epsilon^{ijk} \xi^k.
\ee
The quantities $\chi^{ij}, \xi^i$ then receive the interpretation of
``internal'' stress and momentum correspondingly. However, at this stage, there are no
reasons to require $\chi^{ij}, \xi^i$ to be real, while the
similar quantities in the ``spacetime'' stress-momentum $T^{ij}$ are observable
quantities and thus must be real. We note that the
various pieces in the decomposition (\ref{T-decomp}) have the 
interpretations of stress and momentum (internal and ``spacetime'' ones),
but there is no energy density anymore, the later appearing as the
combination of the traces of the internal and spacetime stress tensors. 
For this reason, and also for brevity, we shall refer to $T^i$ as the stress-momentum 
two-form from now on.

Let us now rewrite our field equations as relations between the curvature
and stress-momentum components:
\be\label{field-eqs}
F^i(A_B)=\Phi^{ij} B^j + 4\pi G T^i \qquad \Longleftrightarrow \qquad M^{ij}=\Phi^{ij}+2\pi G \, Q^{ij},
\quad N^{ij} = -2\pi G \, T^{ij},
\ee
where as before the ``gravitational'' quantity $\Phi^{ij}$ is at this stage arbitrary
and is to be determined via the help of Bianchi identities. In the field equations (\ref{field-eqs}) the
quantity $G$ is the usual Newton's constant. Note then that the quantities $Q^{ij}, T^{ij}$ 
must have the dimensions of energy density, so the above interpretation of the components of $Q^{ij}$ as 
the ``internal'' stress-momentum is consistent with their dimensions.

Now, let us, as before, construct an action that 
leads to (\ref{field-eqs}). We have:
\be\label{full-action}
S[B,A,\Phi,\phi_m]=\int B^i\wedge F^i(A) - \frac{1}{2}\Phi^{ij}
B^i\wedge B^j - 4\pi G S_m[B,\phi_m],
\ee
where $\phi_m$ is a collective notation for all the matter fields, and $S_m[B,\phi_m]$ stands for
the matter part of the action, which is assumed to depend on the ``gravitational'' background only via 
the two-form field $B^i$. In principle, one can also envisage the possibility of the
matter fields (e.g. fermions) coupling directly to the connection $A^i$, but this does not bring
anything conceptually new, only complicates the analysis, so we shall not consider this possibility
any further. Defining:
\be\label{T-def}
T^i := \frac{\delta S_m}{\delta B^i}
\ee
we get the equation (\ref{field-eqs}) when varying the action with respect to the two-form
field $B^i$, and the compatibility equations $D_A B^i=0$ when varying the action with
respect to the connection. 

The next step in interpreting the above theory is to note that the stress-energy-momentum
two-form $T^i$ satisfies some conservation laws. Indeed, since the matter part $S_m$ 
of the action must be diffeomorphism invariant, the following identity must hold:
\be\label{non-vac-1}
0=\delta_\xi S_m= \int \frac{\delta S_m}{\delta \phi_m} \delta_\xi \phi_m + \int T^i \wedge \delta_\xi B^i,
\ee
where $\delta_\xi$ is a variation of fields under an infinitesimal diffeomorphism generated by 
a vector field $\xi$. The first term in (\ref{non-vac-1}) vanishes by matter equations of motion,
while the other term gives:
\be\label{T-diffeo-1}
0= \int T^i \wedge D_{A_B} \iota_\xi B^i,
\ee
where $\iota_\xi$ denotes contraction of a form with a vector field, so e.g. $\iota_\xi B^i$ 
is a one-form with components $(\iota_\xi B^i)_\mu:=\xi^\alpha B_{\alpha\mu}^i$.
This expression follows from the following formula for the action of diffeomorphisms on $B^i$:
\be
{\cal L}_\xi B^i = \iota_\xi D_A B^i + D_A \iota_\xi B^i, 
\ee
where $A$ can be taken to be arbitrary, and the fact that $D_A B^i=0$ for $A=A_B$. Now
integrating (\ref{T-diffeo-1}) by parts and taking into account that $\xi$ may be of
compact support, we can conclude that the integrand must vanish:
\be\label{T-diffeo-2}
D_{A_B} T^i \wedge \iota_\xi B^i =0.
\ee
This should hold for any vector field $\xi$, so we get four ``conservation'' equations.

Another important identity that we can obtain for $T^i$ follows from gauge invariance
of the action. Thus, we similarly write (\ref{non-vac-1}) but now consider the variations
of the fields under a gauge transformation. For the $B^i$ field this is:
\be
\delta_\omega B^i = \omega^{ij} B^j,
\ee
where $\omega^{ij}$ is an infinitesimal anti-symmetric matrix - a Lie-algebra element 
of ${\rm SO}(3)$. We can therefore conclude that
\be
T^i \wedge \omega^{ij} B^j =0
\ee
for any matrix $\omega$ and thus:
\be
T^{[i}\wedge B^{j]} =0.
\ee
In the language of the decomposition (\ref{T-decomp}) this translates into:
\be\label{T-gauge}
Q^{[i|l} h^{l|j]}=0 \Longleftrightarrow \epsilon^{ijk} Q^{jl} h^{lk}=0.
\ee
This identity is satisfied by any $T^i$ that follows via (\ref{T-def}) from
a gauge-invariant action, and will be of importance below.

We can now derive a Bianchi identity for the gravitational quantities $\Phi^{ij}$.
Let us rewrite (\ref{bianchi-1}) as:
\be
D_{A_B} \Phi^{ij} \wedge B^j + 4\pi G D_{A_B} T^i = 0.
\ee
Let us now take the wedge product of this expression with the one-form $\iota_\xi B^i$. Using
(\ref{T-diffeo-2}) we see that the field $\Phi^{ij}$ must satisfy:
\be\label{Phi-ident-*}
\iota_\xi B^i \wedge D_{A_B} \Phi^{ij} \wedge B^j \sim h^{ij}D_{A_B}\Phi^{ij}=0,
\ee
where we have used (\ref{key-ident}).
We can now write down all the terms in the expression for the covariant derivative
$D_{A_B}\Phi^{ij}$. Similarly to what we had in the pure gravity case, there is
a term containing the connection $A_B^i$ and proportional to $\epsilon^{ijk}\Phi^{jl}h^{lk}$.
However, we can again conclude that this term is zero. Indeed, the second Bianchi
identity (\ref{bianchi-2}) together with field equations (\ref{field-eqs}) says:
\be\label{Phi-Q-eps}
\epsilon^{ijk} (\Phi^{jl}+2\pi G \, Q^{jl}) h^{lk} =0.
\ee
However, we have seen above that the invariance of the material action under
gauge transformations implies (\ref{T-gauge}), and thus the second term here is
zero, which implies that the first term is zero as well. Thus, from (\ref{Phi-ident-*})
we conclude that:
\be
h^{ij} d\Phi^{ij}=0,
\ee
which is exactly what we had in the pure gravity case. All the remaining steps
from the previous Section go unchanged: we arrive at conclusion that the
potential term in the action proportional to $V=h^{ij} \Phi^{ij}$ is a 
homogeneous function of order one in ${\rm Tr}(h)$ times the cosmological
constant $\Lambda$ times an arbitrary dimensionless 
function $U(H)$. Thus, the full theory is obtained as simply the gravitational 
plus matter parts:
\be\label{action-full-1}
S[B,A,\phi_m]=\int B^i\wedge F^i(A) - \Lambda \frac{U(H)}{6} B^i\wedge B^i - 4\pi G\, S_m[B,\phi_m].
\ee
This solves the problem of coupling of the class of generalized
gravity theories we have been considering to generalized matter.
It only remains to supplement the matter part of the action with
some appropriate reality conditions, for it is in general complex.
A way to do this is to require the components of $T^i$ that are 
``directly'' observable to be real. However, to be able to do
physics with our gravity theory we need to understand how material
actions $S_m[B,\phi_m]$ can be formed and which stress-momentum two-forms
$T^i$ can arise in our non-metric theory.

\section{Stress-momentum two-form of a ``small'' body}
\label{sec:stress-energy}

In the previous Section we have seen that the matter stress-momentum two-form $T^i$
satisfies the following ``conservation'' equations:
\be\label{T-conserv}
\iota_\xi B^i  \wedge  D_{A_B} T^i =0, \qquad T^{[i}\wedge B^{j]} =0.
\ee
In GR the stress-momentum has the special form (\ref{T-general}) and the second
of these equations is automatically satisfied, while the first gives the
usual conservation of energy equation when the two-form field $B^i$ is metric
$B^i\wedge B^j\sim\delta^{ij}$. The conservation equation can then be used to
conclude that ``small bodies'' move in GR along geodesics. A very important
question for the theory just developed is what the notion of geodesic generalizes
to in the case of an arbitrary two-form field background. To understand this
we shall employ the same methods as are used in GR. 

A particularly efficient method that allows to study this question 
has appeared recently in a paper by Gralla and 
Wald \cite{Gralla:2008fg}. This work employs the machinery of asymptotic expansions
to derive results on motion of ``small'' bodies in GR both in the leading approximation,
which gives the result that bodies move along geodesics, as well as in the
sub-leading one, which leads to results on gravitational self-force. For our
purposes we only need the analog of the first of these. Thus, most of the machinery
developed in this work is actually unnecessary here. However, some key ideas of
Section IV of this paper will still be used.

As in \cite{Gralla:2008fg}, the first step is to derive, using the asymptotic expansion techniques,
that the stress-energy-momentum tensor has a well-defined limit approaching a distribution.
To this end, consider a family $B^i(\lambda, x^\alpha)$ 
of two-form field backgrounds. Here $\lambda\geq0$ parametrizes members of
the family and $x^\alpha$ are coordinates of some convenient coordinate system. 
We shall assume there exist coordinates such that $B^i(\lambda, x^\alpha)$ 
are smooth in both $\lambda$ and $x^\alpha$ at least sufficiently far away
$r> \bar{R} \lambda, r^2=\sum_{i=1}^3(x^i)^2$ from the particle, where
$\bar{R}$ is some universal constant, and that for all $\lambda, r>\bar{R} \lambda$
the two-form $B^i(\lambda,x^\alpha)$ is a solution of vacuum field equations of our theory. 
Let us consider the following expansion for the two-form field as $\lambda\to 0$:
\be\label{lambda-exp}
B^i(\lambda) = 
B^i + \lambda b^i +O(\lambda^2),
\ee
where for now both the background $B^i$ and the ``perturbation'' $b^i$ are
functions of all coordinates $x^\alpha$, the background $B^i$ satisfies the vacuum
field equations $F^i(A_B)=\Phi^{ij} B^j$, and the ``perturbation'' satisfies the
linearized field equation, at least sufficiently far from the particle $r>\epsilon$.
 
Now given a background $B^i$ satisfying the field equations, let us define an 
operator $G^i_{\mu\nu}(b)$ via:
\be
F^i(A_{B+\lambda b}) - \Phi^{ij} (B^j+\lambda b^j) = \lambda G^i(b) + O(\lambda^2).
\ee
In terms of $G^i$ the linearized field equations read $G^i(b)=0$.

By definition, the stress-momentum two-form is defined as a distribution
on spacetime whose action on an arbitrary smooth Lie-algebra valued
anti-symmetric tensor $f^i_{\mu\nu}$ is given by integrating the
right-hand-side of linearized field equations, i.e. $G^i(b)$ against
$f^i$. Or, using the fact that $G^i$ is self-adjoint, we can define:
\be
4\pi G T(f) = \int_M G^i(f) \wedge b^i,
\ee
where $b^i$ is as in (\ref{lambda-exp}). 

The definition of the notion of self-adjoint that is used here is as follows. 
It can be shown that for arbitrary two-forms 
$b^i, f^i$ the four-form:
\be\label{self-adjoint}
b^i \wedge G^i(f) - G^i(b) \wedge f^i = d X(b,f)
\ee
is a total derivative, where $X(b,f)$ is a certain 3-form depending on both $b$ and $f$,
as well as the background $B^i$. 
We shall not attempt demonstrate this property in the present paper, as the computation is
quite technical and it would take us too far. 

If we now integrate the expression (\ref{self-adjoint}) over a region $r> \epsilon$ and use
the fact that $b^i$ satisfies the linearized field equations $G^i(b)=0, r>\epsilon$. We
get:
\be
\int_{r>\epsilon} b^i \wedge G^i(f) = \int_{r=\epsilon} X(b,f).
\ee
Taking the limit $\epsilon\to 0$ we get:
\be
T(f) = \frac{1}{4\pi G} \lim_{\epsilon\to 0} \int_{r=\epsilon} X(b,f).
\ee
Similar to what happens in the GR case, it can be seen that this limit exists and
is different from zero if $b^i\sim 1/r$ as $r\to 0$. Thus, when this is the case,
the stress-momentum distribution can be written as:
\be
T(f) = \int dt \, \epsilon^{\mu\nu\rho\sigma} \, T_{\mu\nu}^i(t) f_{\rho\sigma}^i(t).
\ee
Here $T^i(t)$ is a Lie-algebra valued two-form along the curve $\gamma$ (given in the
chosen coordinate system by $x^i=0$), and $f^i(t):=f^i(t,r=0)$ is the value of the test two-form
$f^i$ along the curve $\gamma$. We note that to write this expression we have chosen
a particular background metric (see below on how this is done). Also note, for future use, that 
under conformal transformations of the background metric 
$dt\to \Omega dt, \epsilon^{\mu\nu\rho\sigma}\to \Omega^{-4} \epsilon^{\mu\nu\rho\sigma}$,
and so for the above distribution to be invariant under conformal transformations
of the auxiliary background metric the stress-momentum two-form must transform as 
$T_{\mu\nu}^i(t)\to \Omega^3 T_{\mu\nu}^i(t)$. We shall use this fact in the next Section 
when we check the behaviour of the evolution equations under conformal transformations.

Now, using the Bianchi identity that holds in our theory, 
one concludes that the above distribution must vanish on test two-forms $f^i$ of the form:
\be\label{test-form}
D_{A_B} \iota_\xi B^i,
\ee
where $\xi$ is an arbitrary vector field and $B^i$ is the background two-form field appearing
in (\ref{lambda-exp}). As in \cite{Gralla:2008fg},
to derive consequences of the arising ``conservation'' equations, we shall first
consider the case of special vector fields of the form:
\be\label{vf-special}
\xi^\mu = x^\alpha F(x^1,x^2,x^3) c^\mu(t), \qquad i=1,2,3,
\ee
where $F(x^1,x^2,x^3)$ is an arbitrary function such that $F(r=0)=1$. As we have already mentioned
the coordinates $x^i$ are chosen in such a way that the curve $\gamma$ along which the body is moving
(i.e. in the neighbourhood of which vacuum field equations are assumed) corresponds to $r=0$, where
as usual $r^2=\sum_{i=1}^3 (x^i)^2$. This still, however, leaves a considerable freedom in
the choice of the coordinates. Let us use the background two-form field $B^i$ to help
with this. Thus, recall that $B^i$ defines a conformal metric. Since any metric is locally flat, as 
in \cite{Gralla:2008fg}, we can always choose the coordinates locally so that this conformal
metric is just the Minkowski metric. This means that, without loss of generality we can assume 
the two-form field $B^i$ in the small neighbourhood of $\gamma$ to be given by:
\be
B^i=\Lambda^{i\ui} \B^\ui,
\ee
where the two-forms $B^\ui$ are those describing the Minkowski spacetime:
\be\label{B-metric}
\B^\ui= idt\wedge dx^\ui - \frac{1}{2}\epsilon^{\ui\uj\uk} dx^\uj\wedge dx^\uk,
\ee
and $\Lambda^{\ui\uj}$ are arbitrary matrix-valued functions of spacetime coordinates. 
Note that we have introduced a new type of indices - underlined ones, to
distinguish between the ``internal'' ${\rm SO}(3)$ bundle where the original
fields take values and the ``metric'' bundle where the metric two-form
field (\ref{B-metric}) lives. The covariant derivative $D_A$ only acts on the
original non-underlined indices. The matrix $\Lambda^{i\ui}$ is defined
modulo conformal rescalings of the metric introduced, which sends 
$B^i_m\to \Omega^2 B^i_m$, and, since the background two-form field is by itself 
independent of any choice of the metric, transforms $\Lambda^{i\ui}\to \Omega^{-2} \Lambda^{i\ui}$. 
One can also do a Lorentz rotation on the metric two-forms (\ref{B-metric}) that
acts on the underlined indices. Thus, the quantities $\Lambda^{i\ui}$ are only defined modulo such 
conformal rescalings and ${\rm SO}(3)$ rotations.

Now using the special vector fields (\ref{vf-special}) in the test two-form (\ref{test-form}),
and taking into account that only the term in which the exterior derivative acts on the
coordinate functions $x^i$ gives a non-zero contribution in the limit $r\to 0$, we 
get that the stress-momentum distribution must vanish on the following set of test two-forms:
\be
\Lambda^{i\ui}(t) dx^\ul  \wedge ( i c^0(t) dx^\ui - ic^\ui(t) dt+ \epsilon^{\ui\uj\uk} c^\uj(t) dx^\uk)
\ee
for any choice of $\ul$ and functions $c^0(t), c^\ui(t)$. This gives us $3\times 4$ conditions
on $T^i(t)$ that we would like to exploit to deduce the form of the stress-momentum. Here and
in what follows the notation $f(t)$ stands for the value of the function $f$ along the
curve $r=0$. Thus, $f(t):=f(t,r=0)$.

It is now convenient to consider a related quantity: $\T^\ui(t) := T^i(t)\Lambda^{i\ui}(t)$. The equation
in question then becomes:
\be
\T^\ui(t) \wedge dx^\ul  \wedge ( i c^0(t) dx^\ui - 
ic^\ui(t) dt- \epsilon^{\ui\uj\uk} c^\uj(t) dx^\uk)=0.
\ee
Let us decompose:
\be
\T^\ui(t) = i A^{\ui\uj}(t) dt\wedge dx^\uj - \frac{1}{2} B^{\ui\uj}(t) 
\epsilon^{\uj\uk\ul} dx^\uk\wedge dx^\ul,
\ee
where $A^{\ui\uj}(t), B^{\ui\uj}(t)$ are some unknown matrix-valued functions of time. Setting
$c^\ui(t)=0$ and thus extracting the $c^0$ component of the conservation equations we immediately get:
\be
A^{[\ui\uj]}=0.
\ee
Setting $c^0(t)=0$ and $c^\ui(t)\sim\delta^{\ui\um}$ we get, after some algebra:
\be
B^{\um\ul} + {\rm Tr}(A)\delta^{\ul\um} - A^{\ul\um}=0,
\ee
where we have suppressed the dependence on $t$ for brevity.
From this equation we immediately conclude that $B^{ij}$ is also a symmetric matrix,
and that ${\rm Tr}(B) = -2 {\rm Tr}(A)$, while the traceless parts of $A^{\ui\uj}, B^{\ui\uj}$
are equal. Let us denote these traceless parts by $\chi^{\ui\uj}(t)$, and (a multiple of) the
trace of say $B^{\ui\uj}$ by $m(t)$. Then we obtain the following form of the stress-momentum
distribution:
\be\label{T-form}
\T^\ui(t) = \frac{1}{6}m(t) B^\ui - \frac{1}{2}m(t) \bar{B}^\ui 
+ \chi^{\ui\uj}(t) B^\uj,
\ee
where we have used the definition (\ref{B-metric}) of the Minkowski space two-forms, and
$\bar{\B}^\ui:=-(\B^\ui)^*$ is the anti-self-dual two-forms. It is instructive to compare this
result to the GR one. In that case $\Lambda^{i\ui}=\delta^{i\ui}$, and no non-trivial self-dual
part of the stress-momentum tensor is possible, so $\chi^{\ui\uj}=0$. The remaining two-form
is that corresponding to the ideal pressureless fluid, see (\ref{T-fluid}), with coordinates
chosen such that the momentum $u^i=0$, as it should. We have thus recovered the GR result,
formula (45) of \cite{Gralla:2008fg}. We see that the main modification arising in
our case is the presence of an arbitrary traceless part $\chi^{\ui\uj}$ in the self-dual
part of the stress-momentum two-form. 

The second ``conservation'' equation in (\ref{T-conserv}) can also be exploited. After some
simple algebra we find that along the curve $r=0$ it is equivalent to the condition:
\be\label{T-cons-2}
\epsilon^{ijk} (\Lambda^{-1})^{i\ui}(t) \chi^{\ui\uj}(t) \Lambda^{j\uj}(t) = 0.
\ee
Let us also note the transformation properties of the quantities that appeared in (\ref{T-form}).
Since $T^i\to \Omega^3 T^i$, we have $m(t)\to \Omega^{-1} m(t), \chi^{\ui\uj}(t)\to \Omega^{-1} \chi^{\ui\uj}(t)$,
which are the correct transformation properties for the quantities having the dimensions of mass. 

\section{Motion of a ``small body''}
\label{sec:motion}

Having extracted the form (\ref{T-form}) of the stress-momentum two-form $T^i$ of a ``small body''
we are ready to find equations that such a body must satisfy during its motion. Thus, we
are looking for an analog of the GR statement that ``small bodies'' move along geodesics.
To this end we once again use the fact that the stress-momentum distribution, whose form
(\ref{T-form}) we have determined above, must vanish on test two-forms of the form 
$D_{A_B} \iota_\xi B^i$, where $\xi$ is an arbitrary vector field. 

The computation one has to do is conceptually clear, but a bit involved. A particularly 
efficient way to do it is to use spinors. However, the resulting
intermediate formulae are not particularly transparent for somebody not familiar
with spinor techniques. For this reason we shall use a shortcut based on the fact that changes to the
final result only come from the self-dual part of the stress-momentum two-form, and the 
contribution of the anti-self-dual sector is completely unchanged from the GR case. This has 
to be verified, and, as we have said, the easiest way to do this is to use spinors. We give a complete
derivation in the Appendix. Here we only deal with (the most interesting) self-dual
part, which can be easily done without spinors.

Thus, we decompose the stress-energy distribution two-form $T^i$ into its
self- and anti-self-dual parts and write:
\be\label{cons-1}
\int dt \left( T^i_{sd}(t) + T^i_{asd}(t)\right) \wedge D_{A_B} \iota_\xi B^i =0.
\ee
This must hold for any vector field $\xi$. 

Let us analyze the $T^i_{sd}(t)$ term of (\ref{cons-1}). In the previous subsection we
have found that this part of $T^i$ can be written as:
\be\label{T-sd}
T^i_{sd}(t) = \frac{1}{2} Q^{ij} B^j,
\ee
where $B^i$ is the background two-form field and $Q^{ij}$ is a tensor given by:
\be
Q^{ij} = (\Lambda^{-1})^{i\ui}(t) 
\left( \frac{1}{3} m(t) \delta^{\ui\uj} + \chi^{\ui\uj}(t) \right) (\Lambda^{-1})^{j\uj},
\ee
where $\chi^{\ui\uj}(t)$ is as introduced above in (\ref{T-form}) and the matrix
$(\Lambda^{-1})^{j\uj}$ on the far right is {\it not} evaluated at $r=0$. We will only be interested in this
tensor along the curve $r=0$. Then the quantity $Q^{ij}(t)$ is symmetric 
$Q^{[ij]}(t)=0$ because $\chi^{\ui\uj}(t)$ is symmetric. We also get the following simple
expression for the ``mass'' $m(t)$:
\be\label{Q-m}
h^{ij}(t) Q^{ij}(t) = m(t),
\ee
where $h^{ij}$ is the ``internal'' metric now defined as:
\be\label{h-metr}
h^{ij} = \Lambda^{i\uk}\Lambda^{j\uk}.
\ee
We also see that the second ``conservation'' equation (\ref{T-cons-2}) becomes in terms of $Q^{ij}$:
\be\label{Q-h-eps}
\epsilon^{ijk} Q^{jl}(t) h^{lk}(t)=0.
\ee

Now using the fact that $D_{A_B} B^i=0$ we can take the two-forms $B^j$ in (\ref{T-sd}) under
the operator of covariant derivative. We then use the identity (\ref{key-ident}) to get for
this first term:
\be\label{sd-1}
\frac{1}{4} \int dt \, Q^{ij}(t) D_{A_B}\iota_\xi (B^i\wedge B^j) \sim 
\int dt\,\left( \frac{1}{4} Q^{ij}(t) (\xi^a D_a h^{ij})(t) 
+ \frac{m(t)}{4} (\nabla_a \xi^a)(t) \right),
\ee
where we have omitted unimportant numerical factors and used the background Minkowski
metric (\ref{B-metric}) to write the result. As usual, the notation $f(t)$ stands for 
$f(t,r=0)$ for any function on spacetime. In (\ref{sd-1}) the quantity
$D_a$ stands for the components of the derivative operator that acts on spacetime
indices as the metric compatible one and on the internal indices as the covariant derivative $D_{A_B}$.
To write the last term in (\ref{sd-1}) we have used the relation (\ref{Q-m}). Further, 
the covariant derivative $D$ was replaced in this term by the usual metric compatible one 
because there are no internal indices to act on. A detailed derivation of (\ref{sd-1})
is given in the Appendix. Now using the property (\ref{Q-h-eps}) we see that
the derivative operator $D_a$ in $Q^{ij} D_a h^{ij}$ can be replaced by the partial
derivative one (in place of which we can also use the metric compatible derivative
operator as there are no spacetime indices in this quantity to act on). So, we
get our final result for the self-dual term:
\be\label{sd-2}
\int dt\,\left( \frac{1}{4} Q^{ij}(t) (\xi^a \nabla_a h^{ij})(t) + \frac{m(t)}{4} (\nabla_a \xi^a)(t) \right).
\ee
Note that in the quantity $\nabla_a h^{ij}$ one first evaluates the derivative and
only then evaluates the result at $r=0$.

Let us now treat the second, $T^i_{asd}$ term in (\ref{cons-1}). As we have already
mentioned, this term is exactly the same as it is in GR. This is demonstrated in
detail in the Appendix. So, we have for it:
\be\label{asd-1}
\int dt\, \left(m(t)\, u_a u_b - \frac{1}{4} m(t)\, g_{ab}\right) (\nabla^a \xi^b)(t).
\ee
Here $u^a$ is the vector tangent to $\gamma$, and the quantity in the brackets is
simply the traceless part of the standard particle stress-energy tensor $T_{ab}=m u_a u_b$.

Now adding (\ref{sd-2}) and (\ref{asd-1}) we get:
\be
\int dt \left( \frac{1}{4} Q^{ij}(t) (\xi^a \nabla_a h^{ij})(t) + (m(t)u_a) (u^b \nabla_b \xi^a)(t)\right)=0,
\ee
where we have rewritten the second term in a suggestive form. Integrating in this 
second term by parts, and using the fact that $\xi^a$ is an arbitrary vector field
we get:
\be\label{evolution*}
u^b \nabla_b (m(t) u_a) = \frac{1}{4} Q^{ij}(t) (\nabla_a h^{ij})(t).
\ee

Equation (\ref{evolution*}) is our main evolution equation for ``small bodies''.
It is instructive to see how the GR case gets reproduced. In that case 
there exists a unique background metric such that $h^{ij}=\delta^{ij}$. The 
right-hand-side of (\ref{evolution*}) then vanishes
and we get the usual $u^b \nabla_b (m(t)u_a)=0$, which implies that $m(t)$ is constant along the 
worldline of the ``small'' body, and that this worldline is a geodesic.

As a check of our result (\ref{evolution*}) we must make sure that it is invariant
under conformal transformations of the metric used two write it. Indeed, in the
theory under study only the conformal class of metrics is well-defined, not the
metric itself. To see that the equation (\ref{evolution*}) is conformally invariant
we recall the transformation properties of all the quantities: $m\to \Omega^{-1} m,
Q^{ij}\to \Omega^3 Q^{ij}, h^{ij}\to \Omega^{-4} h^{ij}$ and finally $u_a \to \Omega u_a$,
the last one following from the normalization condition $g^{ab} u_a u_b=1$. Thus, the
quantity $m u_a$ is conformally invariant, and we only need to worry about the 
change of the metric-compatible derivative operator. For brevity we drop the
argument indicating time dependence in all the formulae and get:
\be
\nabla_b (m u_a) \to \nabla_b (m u_a) - 
m( u_b \nabla_a \ln\Omega + u_a \nabla_b \ln\Omega - g_{ab} u^c \nabla_c \ln\Omega).
\ee
This means that the left hand side of (\ref{evolution*}) transforms as:
\be
u^b \nabla_b (m u_a) \to \Omega^{-1} u^b \nabla_b (m u_a) - \Omega^{-1} m \nabla_a \ln\Omega.
\ee
On the other hand, the right-hand-side of (\ref{evolution*}) transforms as:
\be
- \frac{1}{2} Q^{ij} \nabla_a h^{ij} \to - \frac{1}{2} \Omega^3 Q^{ij} \nabla_a \Omega^{-4} h^{ij}=
\\ \nonumber
 - \frac{1}{2} \Omega^{-1} Q^{ij} \nabla_a h^{ij} + 2 \Omega^{-1} Q^{ij}h^{ij} \nabla_a \ln\Omega
= - \frac{1}{2} \Omega^{-1} Q^{ij} \nabla_a h^{ij}- \Omega^{-1} m \nabla_a \ln\Omega.
\ee
which shows that the equation is indeed conformally invariant.

\section{Interpretation}
\label{sec:interp}

In the previous Section we have obtained the ``small body'' evolution
equation for our gravity theory. The only difference with the GR case
stems from the fact that a ``small body'' is allowed to have a non-trivial ``internal stress''
tensor $Q^{ij}(t)$ which then interacts non-trivially with the ``non-metric'' part of
the background. 

From (\ref{evolution*}) we see that, because of the non-zero right-hand-side, 
the motion does not seem to be geodesic. Interestingly,
when the background is metric, i.e. there exists a metric in which $h^{ij}=\delta^{ij}$,
the evolution is geodesic even in case the body has non-trivial ``internal stress'' $Q^{ij}(t)$.
Thus, it is only when the two-form field background in which the body moves is ``non-metric'' 
that we get (apparent, see below) deviations from geodesic motion. 

To give an interpretation to (\ref{evolution*}), let us multiply this equation by
$u^a$. Due to the normalization condition satisfied by this vector field, we have
$u^a \nabla_b u_a=0$, and so:
\be\label{mass-cons}
u^a \nabla_a m(t) = \frac{1}{4} Q^{ij}(t) (u^a \nabla_a h^{ij})(t).
\ee
Thus, the conservation equation no longer implies that the ``mass'' of the
body is constant along the trajectory, but tells us something different. To
see what, we will need to make an additional assumption about the nature
of the ``internal'' stress-momentum of our body. Thus, we shall assume that 
the tensor $Q^{ij}(t)$ that appears in (\ref{evolution*}), (\ref{mass-cons}) is independent
of the direction of motion of the body. That is, we assume that the self-dual
part of the body's stress-momentum two-form is $(1/2)Q^{ij} B^j$ with the
same $Q^{ij}$ no matter along which trajectory the body travels. Note that in general
relativity this is true, with the self-dual part of $T^i$ being given by
$T^i_{sd}=(m/6) B^i$. Our assumption may be motivated by considering how 
the self-dual part of $T^i$ arises from some matter action via (\ref{T-def}).
Indeed, $T^i_{sd}$ arises from a term of the form $(1/2)\tilde{Q}^{ij} B^i \wedge B^j$,
where $\tilde{Q}$ is some matrix possibly depending on the
``non-metric'' components of $B^i$. This suggests that the self-dual, or ``internal'' part 
of the stress-energy two-form should only depend on the internal composition of the particle, and
not on its motion. 

If one makes this well-motivated assumption, then (\ref{mass-cons}) must hold
for any choice of the vector $u^a$. Thus, the following equation must hold:
\be
dm = \frac{1}{4} Q^{ij} dh^{ij}.
\ee
Let us now recall that we have seen a similar equation before, equation (\ref{dV}), 
in Section \ref{sec:mod} that dealt with the vacuum modified gravity. There it implied
that the tensor $\Phi^{ij}$ that arises in the decomposition of the curvature into
its self- and anti-self-dual parts must be given by a derivative of
some potential function with respect to the internal metric. We see that a similar
relation must be true here:
\be\label{Q-h}
Q^{ij}(t) = 4 \left(\frac{\partial m(h^{ij})}{\partial h^{ij}}\right)(t).
\ee
Thus, the evolution equation implies that for each body there must
exist a function $m(h^{ij})$ of the internal metric, such that the tensor
$Q^{ij}$ describing the self-dual part of the stress-momentum of this body
is given by the partial derivatives of the mass function with respect to
the components of the internal metric. Because of this, we shall no longer write the argument 
indicating the time dependence next to $m$, as we now interpret the mass as a function of $h^{ij}$, 
which later must be evaluated on $h^{ij}(t)$. 

Let us now consider the ``mass'' function $m(h^{ij})$ to be a function of the
trace ${\rm Tr}(h)$ of the internal metric and the traceless part $H^{ij}$. Then,
using (\ref{der-ident}), as well as the fact that $m(h)={\rm Tr}(Qh)$ we see
that the function $m({\rm Tr}(h),H)$ is a homogeneous function of
degree $1/4$ in ${\rm Tr}(h)$. Therefore, we can write:
\be\label{M}
m(h)=\bar{m} \frac{({\rm Tr}(h))^{1/4}}{3} W(H),
\ee
where $W(H)=W(H^{ij})$ is a dimensionless function of the traceless part $H^{ij}$ of
the ``internal'' metric $h^{ij}$ normalized as $W(0)=1$, and $\bar{m}$ is
a quantity with the dimensions of mass. Note that the formula (\ref{M}) is
consistent with the transformation properties $m\to \Omega^{-1} m, h\to \Omega^{-4} h$
of the quantities under the conformal transformations of the background metric.
It should also be compared with an analogous formula (\ref{V}) on the gravity side.

Let us now see what the fact (\ref{Q-h}) implies for the motion of the body. We
can now replace the right-hand-side in (\ref{evolution*}) by $\nabla_a m$ to get:
\be
u^b \nabla_b m(t) u_a = (\nabla_a m)(t).
\ee
Recall that this equation is conformally invariant, with
the mass function transforming as $m\to \Omega^{-1} m$. Note also that the mass
function is now defined not only along the trajectory, but everywhere, and we can use the
conformal freedom in choosing the background metric to select a special metric 
in which $m=\bar{m}$ is a constant. Then in this metric, whose conformal factor is defined by the 
condition
\be\label{geod-cond}
{\rm Tr}(h) (W(H))^4 = 3
\ee
the body moves along spacetime geodesics. Note that when the background is metric $H=0$ the mass
$\bar{m}$ is the usual mass of the particle as we know it in general relativity. So,
similar to what we saw happening in the case of pure gravity, the departure from
the familiar behaviour is parametrized by a single dimensionless function $W(H)$
of the traceless part of the internal metric.

It remains to discuss an interpretation of the function $W(H)$. This function is set
by the coupling of the matter component
in question to the two-form field. In principle, it can be arbitrary, and
moreover different for different matter components. In previous studies,
see e.g. \cite{Krasnov:2008sb}, of the theory it was shown that
the non-metricity $H^{ij}$ of the two-form field background is controlled by the Weyl curvature,
and that we can expect this non-metricity in the Solar system with its weak curvatures 
to be extremely small. This means that even if the function $W(H)$ was
different for different species of elementary particles one would not notice
this in the Solar system. However, having different $W(H)$ for different
types of particles would mean that the weak equivalence principle was violated
in the theory. Indeed, in this case each type of particle would travel along
geodesics of its own metric and the universality of free fall would not hold
(in regions of non-metricity). While this is an interesting theoretical possibility
naturally provided by our theory, it is much safer to require the theory
to respect the weak equivalence principle and postulate that the function
$W(H)$ is universal and same for all particles and composite bodies. In
our further development of the theory we shall assume this to be the case.

A universal function $W(H)$ leads to one further interesting consideration. 
Indeed, the assumption of the universality of $W(H)$ is the
assumption that the dependence of all the mass terms in the matter
Lagrangian on the $H^{ij}$ components of the two-form field is the
same. Let us now imagine that we have coupled matter not to gravity as
in (\ref{action-full-1}), but to the topological BF theory:
\be
S_{top}[B,A,\phi_m] = \int B^i \wedge F^i - 4\pi G\, S_m[B,\phi_m].
\ee
Now assuming all our matter fields $\phi_m$ to be quantum, one may take
the vacuum expectation value of the last material term in the Lagrangian
to get a Lagrangian that only depends on the two-form and the connection
fields:
\be\label{action-av}
S_{eff}[B,A] =  \int B^i \wedge F^i - 4\pi G\, \langle S_m[B,\phi_m] \rangle.
\ee
The ``quantum average'' of the matter action must be an integral of 
a four-form that can only be built from the 4-forms $B^i\wedge B^j$.
However, these 4-forms are proportional to the internal metric $h^{ij}$ and
the volume form $B^i\wedge B^i$. Thus, the four-form in question
must be proportional to the volume form. Further, since all the mass terms in $S_m[B,\phi_m]$ depend non-trivially 
on the components $H^{ij}$ of $B$, the quantum average will depend on the function
$W(H)$. Thus, we see that the quantum average in (\ref{action-av}) must be equal to:
\be
S_{eff}[B,A] =  \int B^i \wedge F^i - \frac{\Lambda_{eff}}{6} U_{eff}(H) (B^i\wedge B^i),
\ee
where $\Lambda_{eff}$ is the effective cosmological constant, and $U_{eff}(H)$ is some effective 
potential normalized so that $U_{eff}(0)=1$. Here both $\Lambda_{eff}$ and $U_{eff}(H)$ depend on 
details of the matter Lagrangian and the form of the coupling $W(H)$. What 
we have reproduced via this heuristic argument is precisely the gravitational action (\ref{action-pot})
with the potential $U_{eff}(H)$ and the cosmological constant $\Lambda_{eff}$. This strongly suggests that
the potential function $U(H)$ appearing in gravity must be related to the
mass function $W(H)$ appearing in the matter sector. However, to find this relation one
must perform a complicated quantum computation. In this paper we shall treat both
functions as phenomenological, but one should keep in mind that it should
be possible to relate them in the final theory. 

\section{Coupling to the general stress-energy tensor}
\label{sec:gen}

Considerations of previous Sections fixed the form of the stress-momentum
two-form of a ``small body''. However, in order to be able to develop physical consequences of our 
gravity theory it is necessary to describe how arbitrary types of matter couple to it. 
Fortunately, the above considerations on the form of the stress-momentum two-form of a ``small body'' 
allow us to describe coupling to generic matter.

Since our theory respects the weak equivalence principle and there is a preferred
metric in which test bodies move along geodesics and any metric is locally flat, 
it is natural to postulate, as in general relativity, that all non-gravitational physics 
in our theory is the same as in flat space. In particular, it is natural to assume that the effect 
of any matter on gravity is still characterized just by the usual stress-energy tensor of matter.
Of course, the coupling of this stress-energy tensor to gravity may and will be 
different in the theory under study. 

The question thus reduces to that of describing how the usual stress-energy tensor
of matter couples to our gravity theory. To answer it, the following
formalism will be useful. As we have already done in Section \ref{sec:stress-energy}, given a 
general two-form field background $B^i$, it will be convenient to introduce a set of ``metric'' 
two-forms. To this end, let us choose a representative in the conformal
class defined by $B^i$, and choose a tetrad $\theta^I, I=0,1,2,3$. From the
tetrad one can construct the two forms $B^{IJ}:=\theta^I\wedge \theta^J$ and
take the self-dual part with respect to the indices $IJ$. Let us refer
to the ${\mathfrak so}(3)$-valued two-forms obtains this way as {\it metric}.
They are the two-forms of Plebanski formulation of general relativity
reviewed in Section \ref{sec:Pleb}. As before we denote these metric
forms by a bold letter. Thus, we get a two-form field $\B^\ui$ satisfying the
metricity condition: $\B^\ui \wedge \B^\uj \sim \delta^{\ui\uj}$,
as well as the reality conditions, see (\ref{B-conds}). As before, we
shall continue to use the underlined indices to refer to quantities taking
values in the ``metric'' ${\rm SO}(3)$ bundle that we have introduced via $\B^\ui$. 
Now, given the metric forms $\B^\ui$ defining the same notion of self-duality on two-forms as $B^i$,
the original two-form field $B^i$ can be represented as a linear combination
of the metric ones:
\be\label{Lambda}
B^i = \Lambda^{i\ui} \B^\ui.
\ee
The quantities $\Lambda^{i\ui}$ are defined up to ${\rm SO}(3)$ rotations
and rescalings of the metric two-forms $\B^\ui$. Thus, the invariant
information contained in them is that in $9-4=5$ components, and we can
parametrize a general two-form background $B^i$ by its metric two-forms
$\B^\ui$ and by the quantities $\Lambda^{i\ui}$, modulo conformal and
${\rm SO}(3)$ transformations. We shall see that this parametrization is
very convenient for practical computations. In particular, the internal metric
$h^{ij}$ is given in terms of the matrices $\Lambda^{i\ui}$ by:
\be\label{h-Lambda}
h^{ij} = \Lambda^{i\uk} \Lambda^{j\uk}.
\ee

Let us now consider the stress-energy two-form.
The usual stress-energy tensor $T_{ab}$ can be decomposed into its trace ${\rm Tr}(T)$
and traceless $T_{ab} - (1/4) g_{ab} {\rm Tr}(T)$ parts. As we have explained above,
we would like the stress-energy two-form $T^i$ of our theory to be constructed from the 
quantities ${\rm Tr}(T)$ and $T_{ab} - (1/4) g_{ab} {\rm Tr}(T)$. It can be expected
that the traceless part will enter into the anti-self-dual part of the stress-momentum
two-form to be constructed, and the trace part will enter into the self-dual part. 
Our experience with the stress-momentum two-form of a small body suggests that
the anti-self-dual part of $T^i$ is essentially unchanged, and is given by:
\be\label{T-asd*}
T^i_{asd} = (\Lambda^{-1})^{i\ui}\T^\ui_{asd},
\ee
where $\Lambda^{i\ui}$ is the matrix introduced in (\ref{Lambda}), and $\T^i_{asd}$ is the
``metric'' anti-self-dual stress-momentum two-form, see Section \ref{sec:Pleb}. Indeed,
we have seen in (\ref{T-form}) that for a small body, the anti-self-dual 
tensor $\Lambda^{i\ui} T^i_{asd}$ has the usual form of one in the metric
theory. We assume that this generalizes to arbitrary matter, and later
check that this choice is consistent with energy conservation. This
solves the problem of coupling the traceless part of $T_{ab}$ to
our gravity theory.

The coupling of the trace part ${\rm Tr}(T)$ is also suggested by what happens
in the ``small body'' case. Thus, we saw that for a ``small body''
\be\label{T-sd*}
T^i_{sd} = \frac{1}{2} Q^{ij} B^j,
\ee
and that the internal stress-momentum tensor $Q^{ij}$ is given by 
a derivative of a ``mass function'' (\ref{M}) with respect to the internal metric. We 
shall keep the same relation in the general case and write:
\be
Q^{ij} = {\rm Tr}(T) \frac{\partial R_m(h)}{\partial h^{ij}},
\ee
where ${\rm Tr}(T)$ is the trace of the stress-energy tensor (with dimensions of
energy density), and the dimensionless function $R_m(h)$ is given by
\be
R_m(h) = \frac{{\rm Tr}(h)}{3} U_m(H),
\ee
where $U_m(H):=(W(H))^4$ is the matter sector potential. Unlike in (\ref{M}), 
which uses $({\rm Tr}(h))^{1/4} W(H)$,
we have now used the fourth power of this combination. This is necessary for
the internal stress-momentum tensor $Q^{ij}$ to be invariant under conformal
transformations of the background metric. With this choice of $T(h)$ we get:
\be\label{Q*}
Q^{ij} = {\rm Tr}(T)\left( \frac{\partial U_m}{\partial H^{ij}} +
\frac{1}{3} \delta^{ij} \tilde{U_m} \right),
\ee 
where, as before, $\tilde{U_m}$ is the Legendre transform (\ref{Legendre}) of 
the matter potential $U_m(H)$. Using the identity similar to (\ref{h-Phi-ident}), it is now 
easy to see that
\be\label{dT}
h^{ij} dQ^{ij} = R_m(h) d{\rm Tr}(T).
\ee
We will need this identity below when we discuss the energy conservation.

The expressions (\ref{T-asd*}), (\ref{T-sd*}) and (\ref{Q*}) determine the coupling
of a general stress-energy tensor $T_{ab}$ to our gravity theory. The only
extra input that needs to be specified on top of what is already present in general relativity is 
two dimensionless potentials $U(H),U_m(H)$ of the traceless part $H^{ij}$ of the internal metric.
The potential $U(H)$ determines the dynamics of the vacuum gravity, and
the material potential $U_m(H)$ is necessary to specify the coupling to matter. 

It remains to check that the coupling specified is consistent with the standard
energy conservation. The only thing that needs to be verified is that there
are no changes in the self-dual part of the stress-energy two-form.
The anti-self dual part does not change. A detailed argument involves spinors and is given in 
the Appendix. In our theory the conservation equation for
$T^i$ is given by (\ref{T-conserv}). Using (\ref{T-sd*}), the self-dual part of this 
equation becomes:
\be
\frac{1}{2} \iota_\xi B^i \wedge D_{A_B} Q^{ij} B^j =
\frac{1}{4} \iota_\xi (B^i\wedge B^j) D_{A_B} Q^{ij},
\ee
where we have again used the identity (\ref{key-ident}). Passing to the
description of the two-form field $B^i$ in terms of the metric two-forms $\B^\ui$,
and taking note of the definition (\ref{h-Lambda}) of the internal
metric we can write this as:
\be
\frac{1}{3} \iota_\xi (\B^\uk \wedge \B^\uk) \frac{1}{4} h^{ij} d Q^{ij}=
\frac{1}{3} \iota_\xi (\B^\uk \wedge \B^\uk) R_m(h) d({\rm Tr}(T)/4).
\ee
where we have used the second equation in (\ref{T-conserv}) to replace
the covariant derivative $D_{A_B}$ by the usual one, and used (\ref{dT})
to arrive at the final expression. The only modification here as compared
to GR is the presence of the factor $R_m(h)$ in this expression. We see, therefore, 
that the conservation equation holds in the metric in which 
\be\label{cond*}
{\rm Tr}(h) U_m(H) = 3,
\ee 
in agreement with our finding (\ref{geod-cond}) in the previous Section. This
establishes that the standard stress-energy tensor $T_{ab}$ 
conservation equation is consistent with the conservation equation
(\ref{T-conserv}) for $T^i$ when the stress-momentum two-form is
constructed from the components of $T_{ab}$ as specified
in equations (\ref{T-asd*}), (\ref{T-sd*}) and (\ref{Q*}). A more
thorough discussion of the energy conservation (including a treatment
of the anti-self-dual part) may be found in the Appendix.

This finishes the formal development of our theory. We now have all the ingredients
to study physics with it, as we know how to describe pure gravity, know how it
is influenced by a general stress-energy tensor, and know how test matter moves in a given gravitational
background. At the end, our gravity theory has all the
standard ingredients of general relativity. The only new quantities that
we have introduced are two dimensionless ``potential'' functions 
$U(H), U_m(H)$ depending on the components $H^{ij}$ of the two-form field. 
Choosing the metric conformal factor so that the condition (\ref{cond*}) is satisfied gives
us completely standard physics for matter fields. The only thing
that changes is the coupling of matter to gravity, as well
as the gravitational dynamics.

\section{Recipe}
\label{sec:recep}

We finish our exposition of the new theory with an ``explicitly metric'' 
formulation that is useful for practical computations. The reader, however,
will not see a metric below, only two-forms $B^\ui$ constructed from the metric,
similar to what happens in Plebanski formulation of general relativity
reviewed in Section \ref{sec:Pleb}. A more conventional formulation of the theory 
that uses the spacetime metric explicitly is also possible, see \cite{Freidel:2008ku}
for a recent treatment. 

As in GR in Plebanski formulation, see Section \ref{sec:Pleb}, one starts with a metric
and the corresponding set of self-dual metric two-forms, which
we denote by $\B^\ui$. One then forms a general linear combination 
of the metric forms:
\be\label{B-Bm}
B^i = \Lambda^{i\ui} \B^\ui,
\ee
introduces the ``internal'' metric
\be
h^{ij} = \Lambda^{i\uk} \Lambda^{j\uk},
\ee
and finds its traceless part:
\be
H^{ij} = \frac{3 h^{ij}}{{\rm Tr}(h)} - \delta^{ij}.
\ee
The theory is specified by two dimensionless potential functions normalized as:
\be
U(H) = 1+ \frac{\alpha}{2} {\rm Tr}(H^2) + O(H^3), \qquad 
U_m(H) = 1+ \frac{\beta}{2} {\rm Tr}(H^2) + O(H^3),
\ee
where $\alpha, \beta$ are dimensionless parameters. It can be seen 
that, apart from those already available in GR, these are the only parameters that are of 
relevance for the linearized theory. The matrices $\Lambda^{i\ui}$ are required to satisfy:
\be
{\rm Tr}(h) U_m(H) = 3.
\ee
After this is done, one finds the two-form field compatible
connection $A_B$ such that $D_{A_B} B^i=0$ (note that the derivative
operator $D_A$ only acts on the non-underlined indices). The 
field equations then read:
\be\label{field-eqs-general}
\Lambda^{i\ui} F^i(A_{\Lambda\B}) = \Lambda \left( \Lambda^{i\ui} \frac{\partial U}{\partial H^{ij}} \Lambda^{j\uj} +
\frac{1}{3} h^{ij} \tilde{U} \right) \B^\uj 
+  2\pi G \T \left( \Lambda^{i\ui} \frac{\partial U_m}{\partial H^{ij}} \Lambda^{j\uj} +
\frac{1}{3} h^{ij} \tilde{U_m} \right) \B^\uj - 
2\pi G \T^{\ui\uj} \bar{\B}^\uj,
\ee
where $\tilde{U}, \tilde{U}_m$ are the Legendre transforms (\ref{Legendre}) of the potentials 
$U(H), U_m(H)$, $\Lambda$ is the cosmological constant, $\T$ is the trace of the standard metric 
stress-energy tensor of matter, $\T^{\ui\uj}$ are the quantities constructed from  
the traceless part of the standard stress-energy tensor, and $\bar{\B}^\ui$
are the anti-self-dual metric two-forms. For example, for the ideal fluid:
\be\label{T-fluid-rec}
\T = (\rho-3P), \qquad \T^{\ui\uj} = (\rho+P) \delta^{ij} + i\epsilon^{ijk} u^k,
\ee
where $\rho, P$ are the energy and pressure densities and $u^i$ is (related to) the momentum vector.
The limit to general relativity is obtained by making the gravitational potential infinitely steep, 
i.e. by sending $\alpha\to\infty, H^{ij}\to 0$ so that the product $\alpha H^{ij}$ remains finite. 
When $H^{ij}=0$ the matrix $\Lambda^{i\ui}$ is an arbitrary ${\rm SO}(3)$ one,
for example the identity matrix, and it is evident that (\ref{field-eqs-general})
reproduces Plebanski equations. The only new ingredient in (\ref{field-eqs-general}),
apart from the usual metric and stress-energy tensors, are the quantities $\Lambda^{i\ui}$
that change the gravitational dynamics and the coupling to matter. They are, however,
non-dynamical, and their only job is to ``twist'' the theory as prescribed by two
potentials $U(H), U_m(H)$.

The discussion of the previous sections has demonstrated that the stress-energy tensor
is still conserved in this theory in the usual way $\nabla^a T_{ab}=0$, where
$\nabla^a$ is the metric-compatible derivative, and that test bodies move along
geodesics. In both cases the relevant metric is the one that is used in the construction of
the metric two-forms $\B^\ui$. One can use the formulae given in this section
as a definition of the theory. A reader who finds this definition a bit contrived should
consult earlier sections for a simpler, but more abstract description. The physical exploration 
of this theory is left to future publications.

For applications it is sometimes more convenient to work not with the internal
metric $h^{ij}$, but with the matrices $\Lambda^{i\ui}$ introduced above in (\ref{B-Bm}).
Thus, let us describe an equivalent formulation of the theory in which the internal metric
never appears and one works directly with the quantities $\Lambda^{i\ui}$.
Again, we start with a metric and the corresponding set of self-dual metric two-forms, which
satisfy $\B^\ui\wedge \B^\uj\sim \delta^{\ui\uj}$. As before,
we form a general linear combination of the metric forms (\ref{B-Bm}).
However, now instead of introducing two potentials $U(H), U_m(H)$, let us
work directly with the combinations: $R(h) = ({\rm Tr}(h)/3) U(H), R_m(h) = ({\rm Tr}(h)/3) U_m(H)$,
which are two (arbitrary) ${\rm SO}(3)$-invariant functions of the internal metric $h^{ij}$ normalized so that 
$R(\delta)=R_m(\delta)=1$ and homogeneous of degree one in the quantity ${\rm Tr}(h)$. 
Let us view these functions as those of $\Lambda^{i\ui}$.
Then they are arbitrary (normalized) functions of matrices $\Lambda^{i\ui}$ that are invariant
under left and right action of ${\rm SO}(3)$, and transform as $R\to \Omega^{-4} R, R_m\to \Omega^{-4} R_m$ 
when $\Lambda^{i\ui}\to \Omega^{-2} \Lambda^{i\ui}$. In order for the metric used to construct $B^\ui$ 
to be the physical one, in which matter moves along geodesics, the quantities $\Lambda^{i\ui}$
are required to satisfy the conditions:
\be\label{Lambda-norm}
R_m(\Lambda)=1.
\ee

As before, one now finds the $B$-compatible
connection $A_B$ such that $D_{A_B} B=0$ (note that the derivative
operator $D_A$ only acts on the non-underlined indices). After this is done, one computes
the curvature of the connection $A_{B}$. From (\ref{field-eqs-general}) we see that the 
right-hand side of the field equations contains matrices of the type
$\partial f(h)/\partial h^{ij}$. Let us note an identity:
\be
\Lambda^{i\ui} \frac{\partial f}{\partial h^{ij}} \Lambda^{j\uj} = \frac{1}{2} \frac{\partial f}{\partial \Lambda^{i\ui}}
\Lambda^{i\uj},
\ee
where $f$ on the right-hand-side is considered to be a function of the matrix $\Lambda^{i\ui}$. Using
this identity we can rewrite the field equations (\ref{field-eqs-general}) in a way that uses directly the
quantities $\Lambda^{i\ui}$. One gets:
\be\label{field-eqs-general-lambda}
\Lambda^{i\ui} F^i(A_{\Lambda\B}) = 
\Lambda \left( \frac{1}{2} \frac{\partial R}{\partial \Lambda^{i\ui}} \Lambda^{i\uj} \right) \B^\uj 
+  2\pi G \T \left( \frac{1}{2} \frac{\partial R_m}{\partial \Lambda^{i\ui}} \Lambda^{i\uj} \right) \B^\uj - 
2\pi G \T^{\ui\uj} \bar{\B}^\uj.
\ee
The interpretation of the quantities $\Lambda$, $\T$ and $\T^{\ui\uj}$ is as before.
Thus, for the ideal fluid we have (\ref{T-fluid-rec}). The formulation that works
directly with the internal metric ``triads'' $\Lambda^{i\ui}$ thus leads to more
compact field equations as compared to (\ref{field-eqs-general}) and may be
preferable for some purposes. The ``twisting'' role of the scalars $\Lambda^{i\ui}$
is particularly clear in the formulation (\ref{field-eqs-general-lambda}).

From the described Plebanski-like formulation (\ref{field-eqs-general-lambda})
it may seem that the obtained field equations have little to do with the objects one usually 
works with in gravity, namely the spacetime metric and the stress-energy tensor of some matter that couples
to this metric. However, we would like to stress that in the final formulation our theory is completely 
standard and works exactly with the same quantities. Thus, we have the physical metric and the matter 
couples to it in a completely standard way. The matter moves along geodesics of this
physical metric and has the usual stress-energy tensor. What is non-standard
is how field equations for this metric are obtained. To this end one
introduces certain extra scalar fields $\Lambda^{i\ui}$ and deforms
Einstein equations in a way that does not generate any kinetic term
for the scalar fields and is consistent with energy conservation. 

Our theory in its final form may be compared to the ``modified source gravity''
of \cite{Carroll:2006jn}, where the authors, following \cite{Flanagan:2003rb},
introduce a scalar field $\psi$ and consider a gravity theory described by the following
simple Lagrangian:
\be\label{mod-source}
\int_M \sqrt{-g} (R - U(\psi)) + S_m[e^{2\psi} g, \phi_m].
\ee
Here $g$ is a dynamical metric, but note that the matter couples
not to $g$ but to a conformally related metric $e^{2\psi} g$ instead.
Importantly, the field $\psi$ is non-dynamical, i.e. does not have
a kinetic term. In vacuum, the theory is just GR with a cosmological
constant. But in general the theory for the physical metric
$e^{2\psi}$ is different from GR and, in particular, the stress-energy
tensor of matter sources the Einstein tensor of $e^{2\psi} g$ in a modified way.

The theory we have considered in this paper is similar to (\ref{mod-source})
in the sense that the coupling of the stress-energy tensor of matter to gravity
is modified. The modification also arises from introducing non-dynamical
scalar fields, even though there is now a multiplet $\Lambda^{i\ui}$ of
them instead of a single one in (\ref{mod-source}). However, 
unlike in the case of (\ref{mod-source}), the pure gravity theory is modified as well,
with this modification being controlled by the potential $R(h)$.
Another important difference is that, unlike in the theory (\ref{mod-source}) 
that modifies the homogeneous isotropic Universe solution,
in our theory the scalars $\Lambda^{i\ui}$ are set to ${\rm SO}(3)$ matrices
in this case by symmetry, so the homogeneous isotropic cosmology is unmodified. 
But the principle according to which (\ref{mod-source}) is constructed
is quite similar to that used in our theory. This is made especially
clear by a recent reformulation \cite{Freidel:2008ku} of the theory
that works directly with the spacetime metric.

Let us finish this section with two more remarks. As we see from the 
field equations (\ref{field-eqs-general-lambda}), in general, the self- and
anti-self-dual parts of the stress-momentum two-form, or, in other words, the trace
and the tracefree parts of the stress-energy tensor of matter tensor appear
on the right-hand side of field equations on a different footing. Indeed, there is an extra matrix multiplying the self-dual
part proportional to $\T$ in (\ref{field-eqs-general-lambda}). An interesting question is
if there is any metric among the conformal class defined by the original two-form
field $B^i$ such that both parts of its stress-energy tensor appear in (\ref{field-eqs-general-lambda})
in the same way. Thus, we are looking for a function $R_m(\Lambda)$ such that
$\partial R_m/\partial \Lambda \sim \Lambda^{-1}$, such that $R_m(\delta)=1$ and
which transforms under $\Lambda\to \Omega^{-2}\Lambda$ as $R_m\to \Omega^{-4} R_m$. 
This function is:
\be\label{R-Urbantke}
R_m(\Lambda)=({\rm det}(\Lambda))^{2/3},
\ee
or, in terms of the internal metric $R_m(h)=({\rm det}(h))^{1/3}$. With this choice
of the matter side potential the field equations take the form:
\be\label{field-eqs-urbantke}
\Lambda^{i\ui} F^i(A_{\Lambda\B}) = 
\Lambda \left( \frac{1}{2} \frac{\partial R}{\partial \Lambda^{i\ui}} \Lambda^{i\uj} \right) \B^\uj 
+  4\pi G \T^\ui,
\ee
with $\T^\ui$ given by its usual expression in Plebanski theory, see e.g. (\ref{T-fluid})
for the case of the ideal fluid. Let us finally note that the condition (\ref{Lambda-norm})
in this case is simply ${\rm det}(h)=1$, which defines the so-called Urbantke
metric \cite{Urbantke:1984eb}. Thus, we can rephrase the above discussion by
saying that the Urbantke metric is distinguished in our theory by the fact
that when matter couples to this metric the field equations take a particularly
simple form (\ref{field-eqs-urbantke}). 
Urbantke metric has recently played a distinguished role in a reformulation of
this theory proposed in \cite{Freidel:2008ku}. However, unlike in this reference, 
instead of fixing the metric to which matter couples from the outset, we
prefer to allow matter to couple to an arbitrary metric in the
conformal class of $B^i$, and control this coupling by the
matter side potential $R_m(\Lambda)$. It can be seen that a non-trivial coupling 
with $R_m(\Lambda)$ different from (\ref{R-Urbantke}) leads to new
interesting physical effects absent in the case (\ref{R-Urbantke}),
which in our opinion serves as a sufficient motivation to allow such a more general coupling.

It is also interesting to note that one can obtain a simple but still non-trivial 
theory with both potentials fixed by taking, in addition to (\ref{R-Urbantke}),
the gravitational potential to be given by the same expression $R(h)=({\rm det}(h))^{1/3}$.
The obtained theory has no adjustable parameters and its field equations read:
\be\label{simple}
\Lambda^{i\ui} F^i(A_{\Lambda\B}) = \frac{1}{3} \Lambda \B^\ui +  4\pi G \T^\ui.
\ee
If not for the presence of the quantities $\Lambda^{i\ui}$ the vacuum $\T^\ui=0$ version of these
equations would be just the constant curvature condition. The presence of the extra scalars
twisting this equation makes them more interesting, and, in particular, makes a non-trivial
spherically-symmetric solution possible. However, in view of the fact
that the modified gravity theory (\ref{simple}) does not have adjustable parameters
it is likely to be in gross conflict with the standard gravity tests. Thus,
the particularly simple version of the theory with both potentials fixes to be 
$R(h)=R_m(h)=({\rm det}(h))^{1/3}$ is likely to be of only academic interest.

\section{Discussion}

With most of the discussion being embedded in the main text we shall only make some
``philosophical'' remarks here. The role of the metric in general relativity is two-fold. First, a metric defines
the spacetime causal structure (lightcones at every point). However, to define the 
causal structure one only needs the conformal metric, i.e. the metric modulo conformal
rescalings. Second, every spacetime point in general relativity is effectively equipped with
a set of rulers and a clock. It is for this purpose of measuring spacetime intervals that one 
needs a metric per se, not just a conformal metric. While the propagation of light is
a very basic process that can arguably make sense to be built into the very definition
of the spacetime structure, the availability of rulers and clocks at every point is
on a very different footing. Indeed, a measurement of distances and time intervals is
a complex physical process that requires in each case a macroscopic physical system
- a solid body or a clock. The very availability of rulers and clocks at every spacetime
point is quite striking, for even the empty space, which is by definition void of
anything material, is endowed in GR with this structure. In our opinion this is a very 
anti-Machian feature of GR: according to Mach's ideas only a relative description
of material bodies in the Universe is possible, and an ``empty'' Universe filled with clocks
should be approached with suspicion. One can argue that the notion of spacetime distance 
is a macroscopic one, and has no place in any reasonable microscopic description. The fact
that macroscopic bodies behave as if to register spacetime intervals needs to be
explained, not postulated. This is not so for the causal structure, as the tiny
quanta of electromagnetic field constantly popping out of the vacuum can be argued to 
define the causal structure of spacetime by their very existence. 

It thus seems reasonable to try to formulate a theory of the gravitational field which
is based on the spacetime conformal structure, not on the spacetime metric. Experimentally 
we only know that the gravitational field is a universal long-range interaction. 
As such it should be possible to think about it as occurring due to exchange of some massless 
particles which can thus have only two possible polarizations. In GR these two polarizations 
arise from the gravitational field of a spacetime metric with its ten components and an additional
4-parameter group of gauge symmetries - diffeomorphisms. However, it seems impossible to build a
similar description on just the conformal structure, as it is specified by 9 components,
which is an odd number, and so no straightforward scheme with gauge symmetries (which reduces
the number of DOF by an even number) can bring 9 total components down to two physical.
A theory of gravity that is based on just the conformal structure would also have 
the problem that there would be no preferred scales in it, so the world described
by it would not be realistic. 

The theory that we have formulated in this paper describes spacetime geometry by specifying
its conformal structure. The way this happens is that in addition to the conformal structure
there are other fields in the theory - other components of the gravitational field. The total
number of ``components'' of our basic two-form field is 18 - an even number and taking into
account all the arising constraints it can be seen that the number of the arising polarizations
of the graviton is still two. The vacuum theory is specified by one arbitrary dimensionless
scalar function $U(H)$ of a traceless symmetric $3\times 3$ matrix $H^{ij}$. When one couples
the theory to matter one has to introduce yet another arbitrary function $U_m(H)$ with
similar properties. It is this matter sector potential function $U_m(H)$ that 
can be shown to supply the conformal factor that defines the spacetime metric
in which test bodies move along geodesics. The limit to general relativity is obtained by setting $H^{ij}=0$. 

As we have discussed, the gravity and material sector functions should be related, since
at least in principle it should be possible to compute the gravity potential $U(H)$
as induced by quantum effects involving matter. Indeed, in Section \ref{sec:interp} we have seen how 
a version of ``induced gravity'' scenario is possible in our setting. With the current
state of the development of the theory such a quantum computation remains beyond
our abilities and there is no choice but to treat the two functions as two independent
phenomenological parameters (or rather two infinite set of parameters) of the theory.
It should be noted, however, that the linearized theory is only sensitive to the
two leading parameters from the arising infinite towers of coefficients. A work analyzing the
effect of modification on the cosmological perturbation theory is currently
in preparation. It should also be noted that the spherically-symmetric solution
of the described gravity theory is known, see e.g. \cite{Krasnov:2008sb}. 
Implications of the modification for the motion of test bodies in the
spherically-symmetric background need to be analyzed in light of findings
of this paper. 

Let us conclude by expressing our amazement at how far it was possible to develop
our modified gravity theory just by following its internal logic. Indeed, the 
theory started its life in \cite{Bengtsson:1990qg} as a rather complicated modification of 
the pure connection formulation of GR. However,
as we have seen from the constructions presented in this paper, the theory
turned out to be a very natural generalization of Plebanski gravity tightly 
constrained by ``Bianchi'' identities, both the vacuum theory and its coupling to matter. 
We have also seen that after specifying the matter sector potential $U_m(H)$
(or, equivalently, $R_m(h)$) and thus specyfying to which metric in the conformal
class defined by $B^i$ the matter couples, the theory takes an entirely standard 
metric form.

It should also be emphasized how striking the results described in 
this paper are from the more familiar perspective of metric-based gravity theories. Indeed, it is 
commonly believed that in order to modify gravity one needs to introduce new
propagating degrees of freedom. In addition, it is often said that the gravitational 
coupling to matter is constrained by the energy conservation, which leaves essentially no
freedom. The modification described in this paper
changes GR without adding to it any extra propagating modes. In addition, a similar modification
of the coupling of the stress-energy tensor to gravity becomes possible,
without any contradiction to energy conservation.

In spite of these exciting results much more remains to be done. In
particular, the question of coupling of fermionic matter directly to
the two-form field remains open. As we have seen in this paper, it
is not necessary to answer this question to obtain physical predictions
of the theory, but it will certainly be an essential question when
the quantization is attempted. We thus hope that the theory that
started its life almost 20 years ago in \cite{Bengtsson:1990qg}
will continue to be a source of interesting results for some more years to come.

\section*{Appendix: Spinor techniques and energy conservation}

Let us, as before, select an arbitrary
metric in the conformal class of metrics determined by $B^i$, construct a tetrad, and then use it
to identify the space of rank 2 mixed primed-unprimed spinors $\lambda_{AA'}$ with spacetime
vectors $\lambda^a$ (and also, using the metric, with spacetime one-forms $\lambda_a$). 
Thus, all spacetime indices are converted into spinor
ones. In these notations our basic two-form field $B^i$ is a self-dual two-form $B^i_{AB}\epsilon_{A'B'}$
(there is no component proportional to $\epsilon_{AB}$ which would correspond to the anti-self-dual part).
Thus, the two-form field $B^i$ becomes described in this language by the metric that it defines, as
well as by the quantity $B^i_{AB}$, symmetric in the unprimed spinor indices $AB$. 
In the GR case $B^i_{AB} B^i_{CD}\sim \epsilon_{A(C} \epsilon_{|B|D)}$,
but in general these quantities are arbitrary. The stress-momentum two-form $T^i$ is described by its
self-dual $T^i_{AB}$ and anti-self-dual $T^i_{A'B'}$ components. 

The conditions (\ref{cons-1}) whose consequences we need to explore in spinor notations become:
\be\label{cons-spinors}
\int dt \left( - T^{i\, AC} D_C^{\, A'} \xi_{A'}^{\, B} B^i_{AB} +
T^{i\,A'C'} D_{C'}^{\, A} \xi_{A'}^{\, B} B^i_{AB} \right)=0,
\ee
where $\xi^{AA'}$ is the spinorial representation of the vector field $\xi^a$, and
$D_{AA'}$ is that of the covariant derivative operator $D_{A_B}$. 

Let us now use the form (\ref{T-form}) of the stress-momentum two form. The anti-self-dual 
component of $T^i$ is given by:
\be\label{T-asd-app}
T^i_{A'B'} = (B^{-1})^i_{AB} \T^{AB}_{\,\,\,A'B'},
\ee
where $T_{ABA'B'}$ has the same form as in GR $\T_{ABA'B'}=m u_{(A|A'|} u_{B)B'}$, 
and is just the traceless part $T_{ab}-(1/4)T g_{ab}$ of the stress-energy tensor $T_{ab}$ of ideal
pressureless fluid $T_{ab} = m u_a u_b, u^a u_a=1$. The quantities 
$B^i_{AB}$ is what $\Lambda^{ij}$ become in the spinor 
notations, and $(B^{-1})^i_{AB}$ is the inverse matrix satisfying:
\be
(B^{-1})^{i\, AB} B^j_{AB} = \delta^{ij}, 
\qquad (B^{-1})^i_{\, AB} B^i_{\, CD}=\epsilon_{A(C} \epsilon_{|B|D)}.
\ee

Let us thus look at the second term in (\ref{cons-spinors}). The compatibility equation $DB^i=0$ takes
in spinor notations the following form:
\be\label{app-comp}
D_{A'}^B B^i_{AB}=0.
\ee
This means that the quantity $B^i_{AB}$ can be taken out of the operator of the covariant derivative:
\be
T^{i\,A'C'} D_{C'}^{\, A} \xi_{A'}^{\, B} B^i_{AB} = T^{i\,A'C'}B^i_{AB} D_{C'}^{\, A} \xi_{A'}^{\, B} 
= T^{i\,A'C'}B^i_{AB} \nabla_{C'}^{\, A} \xi_{A'}^{\, B},
\ee
where the last equality is due to the fact that there are no internal indices in the quantity 
that the covariant derivative operator acts on, and so the derivative operator can be replaced 
by the usual metric-compatible one. We now use the form (\ref{T-asd-app}) of the anti-self-dual part
of the stress-momentum two form to conclude that the second term in (\ref{cons-spinors}) is
given by:
\be
T^{\,\,A'C'}_{AB} \nabla_{C'}^{\, A} \xi_{A'}^{\, B} = 
m u_{(A}^{A'} u_{B)}^{C'} \nabla_{C'}^{\, A} \xi_{A'}^{\, B},
\ee
where we have used the fact that $\T_{ABA'B'}=m u_{(A|A'|} u_{B)B'}$.

We can now substitute (\ref{T-sd}) into the first term self-dual term in the conservation equation 
(\ref{cons-spinors}), and use the compatibility equation to take the quantity $B^{j\, AC}$
under the operator of covariant derivative. The first term becomes:
\be\label{app-1}
- \frac{1}{2} Q^{ij} D_C^{\, A'} \xi_{A'}^{\, B} B^{j\, AC} B^i_{AB}.
\ee
We now use the identity (\ref{key-ident}), which in the spinor notations becomes:
\be\label{app-ident}
B^{(i\,\, A}_{\, E}  B^{j)}_{\, AF} = -\frac{1}{2} \epsilon_{EF} h^{ij},
\ee
to rewrite (\ref{app-1}) as:
\be
-\frac{1}{4} Q^{ij} D_A^{\, A'} \xi_{A'}^{\, A} h^{ij} =
-\frac{1}{4} Q^{ij} \xi_{A'}^{\, A} D_A^{\, A'} h^{ij} - \frac{m}{4} D_A^{\, A'} \xi_{A'}^{\, A},
\ee
where we have used the relation (\ref{Q-m}). Now using the identity (\ref{Q-h-eps}) we see 
that we can replace the covariant derivative in the first term here by the ordinary one.
The covariant derivative in the second term can be replaced by the metric compatible one
as the quantity it acts on does not have internal indices. 

Combining it all together we get for the equation (\ref{cons-spinors}):
\be
0=\int dt\left( -\frac{1}{4} Q^{ij} \xi_{A'}^{\, A} \nabla_A^{\, A'} h^{ij}
+ \frac{m}{4} \epsilon_{AB} \epsilon^{A'B'} \nabla_{A'}^{\, A} \xi_{B'}^{\, B}
+m \, u_{(A}^{A'} u_{B)}^{B'} \nabla_{A'}^{\, A} \xi_{B'}^{\, B}
\right),
\ee
where we have rewritten the second term in a suggestive way. We now use:
\be
u_{(A}^{A'} u_{B)}^{B'} + \frac{1}{4} \epsilon_{AB} \epsilon^{A'B'} = u_{A}^{A'} u_{B}^{B'}
\ee
to get:
\be
0=\int dt\left( -\frac{1}{4} Q^{ij} \xi_{A'}^{\, A} \nabla_A^{\, A'} h^{ij}
+m \, u_{A}^{A'} u_{B}^{B'} \nabla_{A'}^{\, A} \xi_{B'}^{\, B}\right),
\ee
or, in the usual tensorial notations:
\be
0=\int dt\left( \frac{1}{4} Q^{ij} \xi^b \nabla_b h^{ij}
+ m u_b \, (u^a \nabla_a \xi^b) \right) = \int dt \left( \frac{1}{4} Q^{ij} \nabla_b h^{ij}
- u^a \nabla_a (m u_b) \right) \xi^b, 
\ee
where to obtain the second equality we have integrated by parts in the second term. Since
the vector field $\xi^a$ is arbitrary we can conclude that:
\be
u^a \nabla_a (m u_b) = \frac{1}{4} Q^{ij} \nabla_b h^{ij}.
\ee
This finishes our proof of (\ref{evolution*}). 

Let us now consider a proof of energy conservation. For this we take the energy conservation
equation in the form (\ref{T-conserv}), which in spinor notations becomes:
\be
\xi_{A'}^{\, B} B^i_{AB} ( - D^{CA'} T_C^{i\, A} + D^{AC'} T_{C'}^{i\,A'} ) =0. 
\ee
Let us transform the anti-self-dual part first. We can use the compatibility
equation (\ref{app-comp}) to take the quantity $B^i_{AB}$ under the operator
of covariant derivative. We get for this term:
\be
\xi_{A'}^{\, B} D^{AB'} B^i_{AB} T_{B'}^{i\,A'} = \xi_{A'}^{\, B} \nabla^{AB'} \T_{AB\, B'}^{\quad A'},
\ee
which coincides with the usual expression in the metric theory. 

Let us now analyze the self-dual term. Here we can replace the self-dual stress-momentum
two-form by its expression (\ref{T-sd}) in terms of the tensor $Q^{ij}$, and take the
two-form $B^i$ out of the operator of covariant derivative. We get for this first term:
\be
- \frac{1}{2} \xi_{A'}^{\, B} B^i_{AB} B_C^{j\, A}  D^{CA'} Q^{ij},
\ee
where $Q^{ij}$ is now given by:
\be
Q^{ij} = \T \frac{\partial R(h)}{\partial h^{ij}},
\ee
with $\T$ being the trace of the stress-energy tensor $T_{ab}$.
Now using (\ref{app-ident}) we get:
\be
\frac{1}{4} \xi_{A'}^{\, B} h^{ij} D_B^{\,\, A'} Q^{ij} = \xi_{A'}^{\, B} R(h) \nabla_B^{\,\, A'} (\T/4),
\ee
where we have used the identity (\ref{dT}). Combining the two terms above, and writing
the result (up to an overall minus sign) in usual vector notations we get:
\be
R(h) \xi^a \nabla_a (\T/4) + \xi^a \nabla^b (\T_{ab} - (1/4) g_{ab} \T)= 0.
\ee
This coincides with the usual conservation equation $\nabla^a T_{ab}$ when the metric is
chosen so that $R(h)=1$. This finishes our demonstration of the fact that the
usual energy conservation holds in our theory.

\bigskip
\bigskip
{\bf Acknowledgement.} The author was supported by an EPSRC Advanced Fellowship and is
grateful to Steve Carlip and Yuri Shtanov for their suggestion to consider ``small bodies'' 
that resulted in this paper. Special thanks are to Yuri Shtanov for his careful
reading of the manuscript and many important suggestions, as well as to
Laurent Freidel for correspondence.


\begin{thebibliography}{0}

\bibitem{Gralla:2008fg}
  S.~E.~Gralla and R.~M.~Wald,
  arXiv:0806.3293 [gr-qc].

\bibitem{Krasnov:2006du}
  K.~Krasnov,
  arXiv:hep-th/0611182.

\bibitem{Plebanski:1977zz}
  J.~F.~Plebanski,
  J.\ Math.\ Phys.\  {\bf 18}, 2511 (1977).

\bibitem{Krasnov:2008fm}
  K.~Krasnov,
  arXiv:0811.3147 [gr-qc].

\bibitem{Bengtsson:1990qg}
  I.~Bengtsson,
  ``The Cosmological constants,''
  Phys.\ Lett.\  B {\bf 254}, 55 (1991).

\bibitem{Capovilla:1991kx}
  R.~Capovilla, T.~Jacobson and J.~Dell,
  Class.\ Quant.\ Grav.\  {\bf 8} (1991) 59.

\bibitem{Bengtsson:2007zzd}
  I.~Bengtsson,
  ``Note on non-metric gravity,''
  Mod.\ Phys.\ Lett.\  A {\bf 22}, 1643 (2007)
  [arXiv:gr-qc/0703114].

\bibitem{Samuel} T.~ Dray, R.~ Kulkarni and J.~ Samuel, 
``Duality and conformal structure,'' 
Journ. Math. Phys. {\bf 30}, 1306 (1989).

\bibitem{Capovilla:1991qb}
  R.~Capovilla, T.~Jacobson, J.~Dell and L.~Mason,
  Class.\ Quant.\ Grav.\  {\bf 8} (1991) 41.

\bibitem{Krasnov:2008sb}
  K.~Krasnov and Y.~Shtanov,
  arXiv:0805.2668 [gr-qc].

\bibitem{Freidel:2008ku}
  L.~Freidel,
  arXiv:0812.3200 [gr-qc].

\bibitem{Carroll:2006jn}
  S.~M.~Carroll, I.~Sawicki, A.~Silvestri and M.~Trodden,
  New J.\ Phys.\  {\bf 8}, 323 (2006)
  [arXiv:astro-ph/0607458].

\bibitem{Flanagan:2003rb}
  E.~E.~Flanagan,
  Phys.\ Rev.\ Lett.\  {\bf 92}, 071101 (2004)
  [arXiv:astro-ph/0308111].

\bibitem{Urbantke:1984eb}
  H.~Urbantke,
  J.\ Math.\ Phys.\ {\bf 25}, 2321 (1984).





\end{thebibliography}
\end{document}